\documentclass[12pt]{iopart}
\usepackage{iopams,mathrsfs,graphicx,hyperref,color}
\expandafter\let\csname equation*\endcsname\relax
\expandafter\let\csname endequation*\endcsname\relax
\newcommand{\gae}{\lower 2pt \hbox{$\, \buildrel {\scriptstyle >}\over {\scriptstyle
\sim}\,$}}
\newcommand{\lae}{\lower 2pt \hbox{$\, \buildrel {\scriptstyle <}\over {\scriptstyle
\sim}\,$}}
\usepackage{amsmath,relsize}

\def\uarr{\uparrow}
\def\darr{\downarrow}
\def\la{\langle}
\def\ra{\rangle}
\newcommand{\bk}[2]{\langle #1|#2\rangle}   
\def\i{\ket{\psi (0)}}

\def\bra#1{\mathinner{\langle{#1}|}}
\def\ket#1{\mathinner{|{#1}\rangle}}
\newcommand{\ee}{\mathrm{e}}
\newcommand{\ii}{\mathrm{i}}

\newcommand{\im}{\mathrm{i}}

\begin{document}
\title{Quantum random walk and tight-binding model subject to projective measurements at random
times}
\author{Debraj Das}
\address{Department of Engineering Mathematics, University of
Bristol, Bristol BS8 1TW, United Kingdom}
\ead{debraj.das@bristol.ac.uk}
\author{Shamik Gupta}
\address{Department of Physics, Ramakrishna Mission Vivekananda
Educational and Research Institute, Belur Math, Howrah 711202, India}
\ead{shamikg1@gmail.com}
\begin{abstract}
        What happens when a quantum system undergoing unitary evolution in time is subject
to repeated projective measurements to the initial state at random times? A question of general
interest is: How does the survival probability $S_m$, namely, the
probability that an initial state survives even after $m$ number of
measurements, behave as a
function of $m$? We address these issues in the context of two paradigmatic quantum systems, one, the quantum random walk
        evolving in discrete time, and the other, the tight-binding model evolving in continuous time,  with both defined on a one-dimensional periodic lattice with a finite number of sites $N$.  For these two models,  we present several numerical and analytical results that hint at the curious
nature of quantum measurement dynamics. In particular, we unveil that when
evolution after every projective measurement continues with the
projected component of the instantaneous state, the average and the typical survival probability decay as an exponential in
$m$ for large $m$.  By contrast, if the evolution continues with
        the leftover component, namely, what remains of the
        instantaneous state after a measurement has been performed, the survival probability exhibits two characteristic $m$ values, namely, $m_1^\star(N) \sim N$ and $m_2^\star(N) \sim N^\delta$ with $\delta >1$. These scales are such that (i) for $m$ large and satisfying $m < m_1^\star(N)$, the decay of the survival probability is as $m^{-2}$,  (ii) for $m$ satisfying $m_1^\star(N) \ll m <m_2^\star(N)$, the decay is as $m^{-3/2
        }$,  while (iii) for $m \gg m_2^\star(N)$, the decay is as an exponential.  The results for the dynamics with the leftover component, already known for the case of measurements at regular intervals, are being extended here to the case of measurements at random intervals.  We find that our results 
hold independently of the choice of the distribution of times between
successive measurements,  as have been corroborated by our results for a wide range of distributions including exponential and power-law distributions as well as for the case of measurements at regular intervals. This fact hints at robustness and ubiquity of our derived results. 
\end{abstract}
\maketitle

\section{Introduction}
\label{sec:Introduction}

Consider a quantum system described by a time-independent Hamiltonian $H$.  Its state at any time $t$ is characterized by a state vector
$\ket{\psi(t)}$ defined in the Hilbert space $\mathcal{H}_S$ of the system $(S)$. The state vector undergoes unitary evolution in time as
$\ket{\psi(t)}=U(t,t_0)\ket{\psi(t_0)};~t>t_0$, where $U(t,t_0)\equiv
\exp(-\im H(t-t_0))$ is the time-evolution
operator~\cite{qua-note-hbar}. Consider
next a
series of instantaneous measurements performed on the system at random
times. Following the measurement postulate of quantum mechanics~\cite{Cohen}, each
measurement involves projecting the instantaneous state of the
system onto the Hilbert space $\mathcal{H}_D \subset \mathcal{H}_S$ of
the measuring device ($D$).
Starting at time $t=0$ with a state vector with unit norm, it is evident
that each projective measurement reduces the norm, and we may ask for its magnitude after a certain number $m \in \mathbb{Z}^+$ of measurements
have been performed on the system.  The norm of the state vector at any time $t>0$ has the physical
interpretation of survival probability of the initial state at that time
instant. To see this,  consider a quantum
particle undergoing motion in a spatial domain
$\mathcal{D}$.  It then follows that
$|\psi({\bf r},t)|^2 {\rm d}^3{\bf r}$ gives the probability of finding the particle between locations ${\bf r}$ and ${\bf r}+{\rm d}{\bf r}$ in $\mathcal{D}$ at
time $t$, with $\psi({\bf r},t)\equiv \bra{\bf r}\psi(t)\ra$. With
$\int_{\mathcal{D}}
{\rm d}^3{\bf r}~|\psi({\bf r},t=0)|^2=1$ so that the particle is
initially for sure within $\mathcal{D}$, and with $m$
instantaneous projective measurements
performed on the system at random times $t_1,t_2,t_3,\ldots,t_m$, with $0<t_1<t_2<t_3<\ldots<t_m$, the norm of the state vector at time $t_m$ given by $\langle \psi(t_m)|\psi(t_m)\rangle=
\int_{\mathcal{D}}
{\rm d}^3{\bf r}~|\psi({\bf r},t_m)|^2$ gives the
probability $S_m$ that the particle is still inside $\mathcal{D}$
at the end of $m$ projective measurements. In other words, $S_m$ is the survival probability that the
particle has survived in $\mathcal{D}$ up to time $t_m$.  With the definition $t_0=0$,  note that $S_0=1$.  For latter purpose,  let us define the quantity 
\begin{align}
F_m \equiv S_{m-1}-S_m\,,
\label{eq:Fm-definition}
\end{align}
which stands for the first-detection probability, namely, the probability that the particle gets detected by the measuring device at time $t_m$ \textit{for the first time}.  It is pertinent to ask: How does $S_m$
vary from one realization of random times $\{t_1,t_2,\ldots,t_m\}$ to
another? What is the dependence on $m$ of $\overline{S_m}$, the averaged survival
probability? How does the typical value of the survival probability, as measured in a typical realization of the random times, depend on $m$? 

\begin{figure}[!htbp]
\centering
\includegraphics[scale=1]{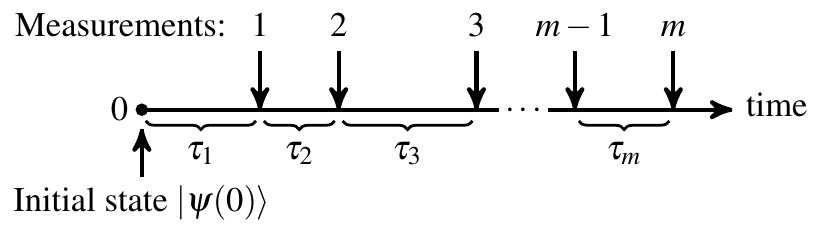}
        \caption{A typical time evolution of a quantum system subject to projective measurements at random times, as
detailed in the text. Starting from the state $|\psi(0)\ra$, the evolution involves
the following repetitive sequence of events: unitary evolution for time according
to operator $U_\alpha$ for time $\tau_\alpha$
        followed by a projective measurement (denoted by the down
        arrows).  In
        \textbf{Scheme 1} (respectively, \textbf{Scheme 2}), subsequent
        evolution at the end of each measurement is carried out with the
        projected component of the instantaneous state (respectively, the leftover component). }
\label{fig:qua-evolution-schematic}
\end{figure}

In the aforementioned protocol,  we may consider the successive time
gaps $\tau_\alpha \equiv
t_\alpha-t_{\alpha-1};~\alpha=1,2,3,\ldots,m$ to
be random variables sampled independently from a common
distribution $p(\tau)$, that is to say, the $\tau_\alpha$'s are independent
and identically-distributed (i.i.d.) random variables.  In this backdrop, let us be more  specific with the framework of our study.  Consider a generic state $\ket{\psi(0)}$ evolving for
a random time $\tau_1$ according to the unitary operator
$U_{1} \equiv \exp{(-{\rm i}H \tau_1)}$.
The evolved state $U_1\ket{\psi(0)}$ is then subject to an
instantaneous projective measurement according to a given projection operator $P$.
Subsequent evolution may then proceed with either the projected
component $PU_{1}|\psi(0)\rangle$ of the state
$U_1|\psi(0)\rangle$ or its leftover component
given by $(\mathbb{I}-P)U_{1}|\psi(0)\rangle$, where $\mathbb{I}$ is the identity
operator. We then iterate $m$ times the aforementioned set of events, so we have two different schemes of time
evolution of the system involving one of the following repetitive
sequence of event-pair:
\begin{itemize}
        \item \textbf{Scheme 1}: A unitary
        evolution for a random time $\tau_\alpha$ according to the unitary
operator $U_\alpha \equiv \exp{(-{\rm i}H\tau_\alpha)}$, followed by the action of the
operator $P$. This scheme corresponds to subsequent evolution with the
                projected component after each measurement.
                \item \textbf{Scheme 2}: A unitary
        evolution for a random time $\tau_\alpha$ according to the unitary
operator $U_\alpha$, followed by the action of the
operator $\widetilde{P} \equiv \mathbb{I}-P$. This scheme corresponds to subsequent
                evolution with the leftover component after each measurement.
\end{itemize}
A typical time evolution for a given realization $\{\tau_\alpha\}_{1\le
\alpha\le m}$ is shown in Fig.~\ref{fig:qua-evolution-schematic}. 
We will take the projection
operator to be performing projection to the initial state:
\begin{align}
P=|\psi(0)\rangle\langle \psi(0)|.  \label{eq:qua-P-operator-def}
\end{align}

In the following, we will use the notation $\ket{\psi_\alpha^{({\rm b})}}$ (respectively,  $\ket{\psi_\alpha^{({\rm a})}}$) to
denote the state of the system at the end of evolution for time
$\tau_\alpha$
and just before (respectively, just after) the $\alpha$-th projective measurement.  It
is evident from the dynamical rules of evolution that
$\ket{\psi_\alpha^{({\rm b})}}=U_\alpha\ket{\psi_{\alpha-1}^{({\rm a})}}$, while
$\ket{\psi_\alpha^{({\rm a})}}$ will be
either $\ket{\psi_\alpha^{({\rm a})}}=P\ket{\psi_\alpha^{({\rm b})}}$
(\textbf{Scheme 1}) or
$\ket{\psi_\alpha^{({\rm a})}}=\widetilde{P}\ket{\psi_\alpha^{({\rm b})}}$ (\textbf{Scheme 2}), and $\alpha=1,2,\ldots,m$.  We define the random variable
\begin{align}
        S_m \equiv S_m(\{\tau_\alpha\}_{1\le \alpha \le m})=\langle \psi_m^{({\rm
        a})}|\psi_m^{({\rm a})}\rangle
        \label{eq:qua-Sm-definition}
\end{align}
as the survival probability of the initial
state after $m$ projective measurements and for the realization
$\{\tau_\alpha\}_{1\le \alpha \le m}$. Note that different values of
$S_m$ correspond to different total duration of evolution $
{\mathcal T}\equiv\sum_{\alpha=1}^m \tau_\alpha$. We may then ask: How does $\overline{S}_m$ depend on $m?$ How does the dependence vary between  \textbf{Scheme 1} and \textbf{Scheme 2}? What is the essential role
played by $p(\tau)$ in dictating the value
of $\overline{S}_m$ in the two schemes? 

\begin{figure}
\centering
\includegraphics[scale=1.08]{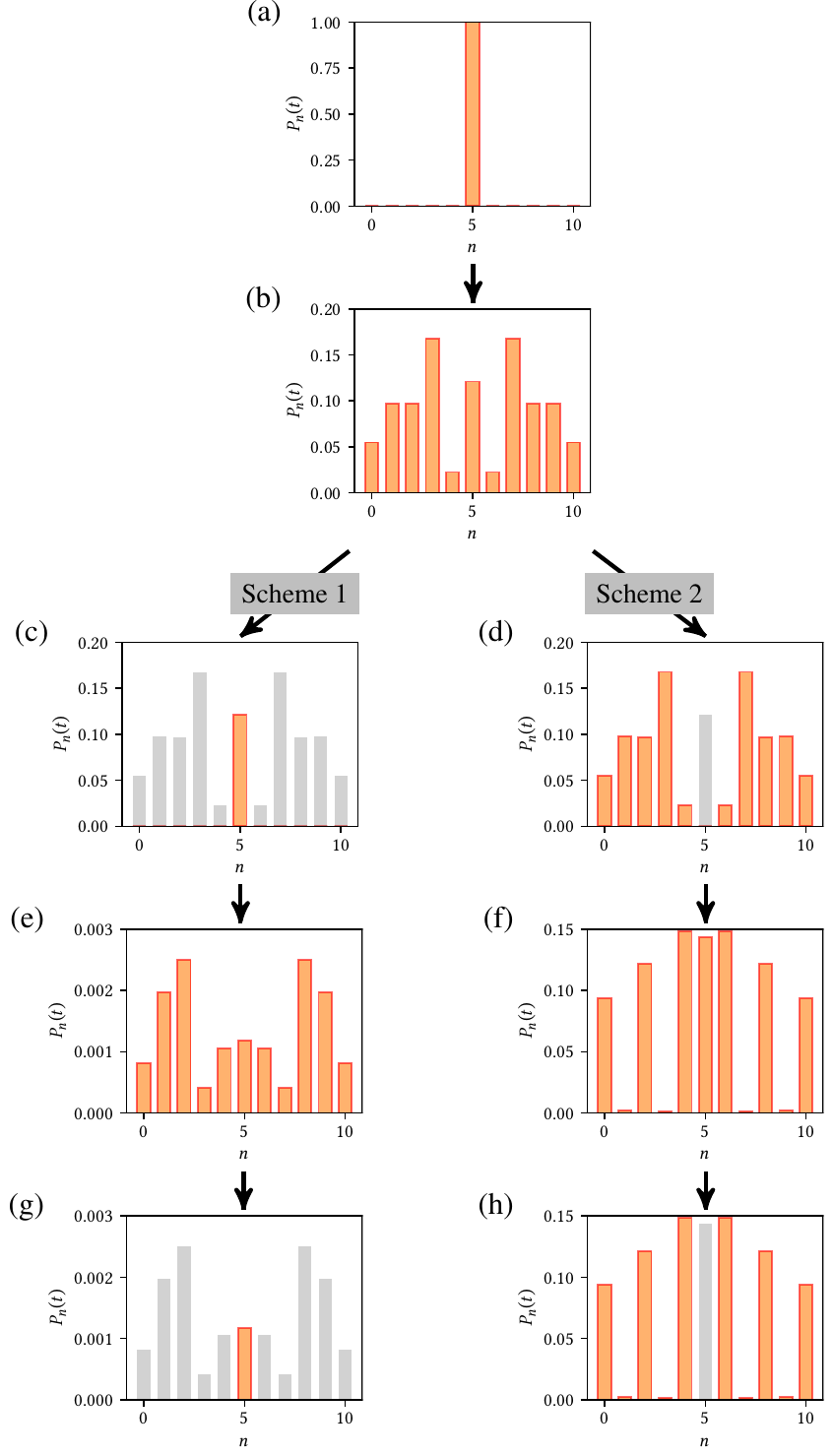}
\caption{Schematic representation of the dynamics studied in the paper under {\bf Scheme 1} and {\bf Scheme 2}.  For a quantum particle moving between the sites of a one-dimensional lattice,  the different panels show the site-occupation probability at different times: panel (a) corresponds to the initial instant $t=0$, panel (b) corresponds to the situation just before the first projective measurement is performed on the system at time $t=t_1$. Panels (c) and (d) denote $P_n(t)$ corresponding respectively to the projected and the leftover state just following the first measurement.  In {\bf Scheme 1} (respectively, {\bf Scheme 2}), subsequent evolution following each projective measurement is carried out with the projected state (respectively, the leftover state). Correspondingly, panels (e) and (f) represent the state just before the second projective measurement performed at time $t_2$ and is the result of unitary evolution of the state represented by panels (c) and (d), respectively.  Finally, panels (g) and (h) represent respectively the projected and the leftover state corresponding to panels (e) and (f), respectively.  The grey bars in these panels show the part of the site-occupation probability that has been projected out by the process of measurement. }
\label{fig:dynamics-schematic}
\end{figure}  

To illustrate with an example the two schemes of dynamical evolution, consider the representative example of a quantum particle moving between the sites of a one-dimensional lattice.  Referring to Fig.  \ref{fig:dynamics-schematic}, wherein $n$ labels the lattice sites, the particle starts its journey at time $t=0$ from site $n=5$ in a lattice of $11$ sites \cite{note-TBM}.  With the ket vector $\ket{n}$ denoting the state of the particle on site $n$ and satisfying $\langle n|m\rangle=\delta_{nm}$ and $\sum_n |n\rangle \langle n|=\mathbb{I}$,  the initial state vector is $\ket{\psi(0)}=\ket{5}$.  In Fig.  \ref{fig:dynamics-schematic}, the different panels show the site occupation probability $P_n(t)=|\langle n|\psi(t)\rangle|^2$, where $|\psi(t)\rangle$ is the state vector at time $t$.  Panel (a) depicts $P_n(t=0)=\delta_{n,5}$. The first projective measurement is performed on the system at time $t_1>0$.  Panel (b) shows $P_n(t)$ just before the first measurement and is the result of unitary evolution of the state in panel (a),  while panels (c) and (d) show respectively results corresponding to the projected and the leftover state just after the first measurement, the state being denoted by the generic symbol $\ket{\psi_{1}^{({\rm a})}}$.  The grey bars in these panels show the part of $P_n(t_1)$ that has been projected out by the process of measurement.   In {\bf Scheme 1} (respectively, {\bf Scheme 2}),  subsequent unitary evolution proceeds with the state corresponding to panel (c) (respectively, panel (d)).  If the second measurement happens at time $t_2>t_1$,  the state $\ket{\psi_{2}^{({\rm b})}}$ corresponds to panels (e) and (f) in the two schemes and is the outcome of unitary evolution of panels (c) and (d), respectively. Panels (g) and (h) show the results for the state $\ket{\psi_{2}^{({\rm a})}}$ just after the second measurement and represent similar to panels (c) and (d) the projected and the leftover state, respectively.  One goes on repeating the above sequence of event-pair: unitary evolution, subsequent evolution with either the projected state ({\bf Scheme 1}) or the leftover state ({\bf Scheme 2}).  In terms of $|\psi_m^{({\rm a})}\rangle$, the site occupation probability just after the $m$-th measurement reads 
\begin{align}
P_n(t=t_m^{({\rm a})})=|\langle n|\psi_m^{({\rm a})}\rangle|^2,
\end{align}
with $t_m^{({\rm a})}$ denoting the time instant just after the $m$-th measurement.
On the other hand,  the survival probability in the two schemes is related to the respective $P_n(t)$ as
\begin{align}
S_m=\sum_n P_{n}(t=t_m^{({\rm a})}).
\end{align}
Thus, the survival probability is nothing but a measure of the site-averaged occupation probability of the particle in the two schemes.
 
The so-called quantum Zeno effect, first discussed in a seminal paper by
Sudarshan and Misra in 1977~\cite{Misra:1977}, is a curious quantum mechanical phenomenon
involving a quantum system subject to projective measurements to a given
initial state at regular time intervals (stroboscopic measurements). In the extreme case of a
frequent-enough series of measurements over a fixed total duration
(implying thereby that the successive measurements are infinitesimally
close to one another), the survival probability for the system to remain
in the initial state may be shown to approach unity in the limit of an
infinite number of measurements, i.e., the dynamical evolution of the
system gets completely frozen in time. Subsequently, stochastic quantum
Zeno effect was
introduced to study the situation in which the measurements are randomly
spaced in time~\cite{Gherardini:2016}. It is important to remark in the context of the present
work that the Zeno effect involves evolution following each measurement
to be carried out with the projected component of the instantaneous
state, while it is evidently of interest to
investigate the dynamics in the complimentary case in which it is the
leftover component of the instantaneous state that undergoes
evolution subsequent to each measurement. These two dynamical scenarios
correspond respectively to {\bf Scheme 1} dynamics and {\bf Scheme 2}
dynamics defined above. Performing measurements at random times is not
quite an
issue of only theoretical curiosity, but may be motivated on the ground that
after all, any experiment that aims to employ projective measurements to
demonstrate the Zeno effect would typically use a timer to time the gap
between successive measurements. Because the timer would invariably be
of finite precision, it would not be possible to ensure that
measurements are performed at exactly regular time
intervals. On the
other hand, in the context of this paper, it is relatively easy to
control the number of times that the projective measurement is repeated,
thus justifying our dynamical set-ups. The average time between two
intervals (averaged over different realizations of the intervals) would of course be finite in experiments.   

In
this work,  we consider two
representative quantum systems defined on a one-dimensional periodic lattice with a finite number of sites $N$,  (i) the quantum
random walk (QRW) model evolving in discrete
times~\cite{qua-Aharonov:1993}, and (ii) the
tight-binding model (TBM), evolving in continuous
time~\cite{qua-Dunlap:1986,qua-Dunlap:1988}.  Within the ambit of these two systems,  we
unveil a plethora of interesting results, numerical as well as analytical,
including universal features in the late-time behavior of both the
average and the typical survival probability, all
of which point to the intriguing nature of quantum measurement process.  We note that experimentally,  it has been possible to implement a random or a stochastic sequence of measurement protocol~\cite{Gherardini:2016,Muller:2017,Gherardini:2017,Gherardini:2018,Do:2019}, and hence we believe that our results are amenable to experimental realization. 

The paper is organized as follows. We choose to study first the QRW,
which is described in detail in Section~\ref{sec:QRW1}.  Here, we compute analytically the site occupation probability, the results of which form a crucial input for the analytical and semi-analytical description of the survival probability pursued in Section~\ref{sec:QRW3}.  This latter section contains our main
results on the survival probability of a generic initial state subject
to instantaneous projective measurements to the initial state at random times. We report extensive
numerical results demonstrating that the
average as well as the typical
survival probability decays asymptotically as an exponential in $m$ when the evolution following each projective measurement is carried on with the projected component of the instantaneous state.  On the contrary, when the evolution proceeds with the leftover component,  the survival probability exhibits two characteristic $m$ values, namely, $m_1^\star(N) \sim N$ and $m_2^\star(N) \sim N^\delta$ with $\delta >1$.  We show that (i) for $m$ large and satisfying $m < m_1^\star(N)$, the decay of the survival probability is as $m^{-2}$ (correspondingly, $F_m$ decays as $m^{-3}$),  (ii) for $m$ satisfying $m_1^\star(N) \ll m <m_2^\star(N)$, the decay is as $m^{-3/2
        }$ (correspondingly, $F_m$ decays as $m^{-5/2}$),  while (iii) for $m \gg m_2^\star(N)$, the decay is as an exponential.  These results
hold independently of the choice of the distribution of times between
successive measurements,  as we demonstrate by our results for a wide range of distributions including exponential and power-law distributions as well as for the case of measurements at regular intervals.  For the projected case, we support our numerical findings with
explicit analytical calculations using large deviation theory (LDT) well
known in probability theory~\cite{qua-ellis-book,qua-Touchette:2009},
while for the leftover case, a semi-analytical approach reproduces correctly
our numerical findings. 

Next, we turn to a description of the TBM and a discussion on its site occupation probability in
Section~\ref{sec:TBM1}. This is followed in Section~\ref{sec:TBM2}\, by a
discussion of our numerical results for the survival probability 
for the case in which the evolution following each projective measurement is
carried on with the projected component as well as for the case in which it
is the leftover component of the instantaneous state that undergoes subsequent evolution until next measurement.
Here too we support our findings with analytical
results derived using the LDT for the projected case and with
semi-analytic calculations for the leftover case. What form the heart of such calculations are our results on site occupation probability presented in Section~\ref{sec:TBM1}.  Similar to the case
of the QRW, we find that for representative $p(\tau)$, the average as well as the typical survival
probability decays as a function of
$m$ as an exponential for the projected case and is characterized by the two  scales $m_1^\star(N)$ and $m_2^\star(N)$ with associated behaviours same as discussed above for the leftover case in the case of the QRW.  Our results for the QRW and the TBM  hint at the robustness of our results with respect to both
discrete and continuous time evolution.  The paper ends with conclusions
in Section~\ref{sec:conclusions}. The appendix collects some of the
technical details of our analytical calculations.

The TBM and related systems when subject to projective measurements with time evolution following {\bf Scheme 2} of dynamics have in recent years been studied to address the issue of when does a quantum particle evolving under the dynamics arrive at a chosen set of sites
\cite{Dhar:2015,Dhar:2015-1,Friedman:2017,Friedman:2017-1,Thiel:2018,Thiel:2019,Lahiri:2019,Meidan:2019,Yin:2019,Thiel:2020,Thiel:2020-1,Dubey:2021,Thiel:2021,Liu:2021,Kessler:2021}.  A crucial difference is that, barring \cite{Kessler:2021}, these contributions
almost exclusively implement measurements at regular intervals of times,  unlike the
scheme considered in this work in which sequence of measurements at random times is implemented.  While we will in later part of the paper discuss in some detail the contribution of this set of work,  let us state right at the outset what new findings we report on with respect to  these references.  Both the quantum random walk and the tight-binding model have been extensively studied in the literature, and the question of first-detection probability has also been amply studied in the past.  However, to the best of our knowledge,  the issue of first-detection probability when measurements are done not at regular but at random time intervals and the ensuing results that we report in this work have not been reported so extensively in the literature.  The fact that the survival probability $S_m$ exhibits in appropriate regimes the decay as $m^{-2}$ and as $m^{-3/2}$, that the first-detection probability $F_m$ decays in appropriate regimes as $m^{-3}$ and as $m^{-5/2}$ with oscillations at small $m$ have all been known in the literature for the case of measurements at regular intervals (see, e.g.,  Refs.  \cite{Dhar:2015,Friedman:2017-1}), and our main contribution as regards {\bf Scheme 2} of the dynamics is to show that these results hold also when measurements are performed at random intervals,  based on our analysis for the studied distributions of these random intervals. 

\section{Quantum random walk (QRW) model}
\label{sec:QRW}

\subsection{Model and site occupation probability}
\label{sec:QRW1}
A quantum random
walk (QRW)~\cite{qua-Aharonov:1993,qua-Vishwanath:2000,qua-Kempe:2003}
is a quantum
mechanical system evolving in discrete times, and involves a walker
performing random walk on a lattice. We consider here the model on a one-dimensional periodic lattice of $N$
sites. Let us label the sites by the index
$n \in \mathbb{Z}$. The motion of the walker on the
lattice depends on the state of a ``quantum'' coin that it carries. The
Hilbert space ${\cal H}_S$ of the walker system is given by a direct sum of the Hilbert space
${\cal H}_P$ for the position of the walker on the lattice and spanned by the
orthonormal basis states $\{|n\ra\};~\la m|n\ra=\delta_{m,n}$, and of the
Hilbert space ${\cal H}_C$ for the coin and spanned by the
orthonormal basis states $|\uarr\ra$ and $|\darr \ra$ representing respectively
the head state and the tail state of the coin. The walker state at any time
$t$, denoted by $|\psi(t)\ra$, is obtained as a linear combination of
the basis states of the space ${\cal H}_S$ with time-dependent expansion
coefficients, where the basis states are given by a direct
product of those of the position and the
coin space. The coin space may also be thought of as a spin space of a
spin-$1/2$ particle, with the basis states being the
eigenstates, the up (u) and the down (d) state, of the $z$-component of
the spin operator. The up and down states may then be taken to correspond respectively to the head and the tail state of the coin.

The random walker evolves according to the following
repetitive sequence of events: acting on the walker state, first by a
unitary operator $ C \otimes \mathbb{I}$, and then by a unitary operator $U$, with both
operations constituting one time step of evolution of the walker state.
Here, $U$ is given by 
\begin{align}
U\equiv|\uarr\ra \la \uarr | \otimes \sum_n |n+1\ra \la n|+|\darr\ra \la \darr |
\otimes \sum_n |n-1\ra \la n|,
\label{eq:qua-U-definition}
\end{align}
while $C$ has the form
\begin{align}
        C \equiv \cos \theta|\uarr\ra\la\uarr| +\sin \theta|\uarr\ra\la\darr| -\sin\theta|\darr\ra\la\uarr|+\cos \theta|\darr\ra\la\darr| ,
\label{eq:qua-Hadamard-definition}
\end{align}
where $\theta \in [0,2\pi)$ is a given parameter. The operator $U$ generates a translation of the
walker on the lattice that is conditioned on the state of the quantum
coin, while the operator $C$ generates a rotation in the coin space.

A large class of
initial conditions corresponds to having the walker on a given site
$n_0$ and
in an arbitrary superposition of the coin states: 
\begin{align}
|\psi(0)\ra=\frac{1}{\sqrt{|a|^2+|b|^2}}\left(a|\uarr\ra+b|\darr\ra\right)\otimes
|n_0\ra,
\label{eq:qua-psi0}
\end{align}
with $a,b \in \mathbb{C}$. Following the rules of evolution of the random walker detailed above, we
may write down explicitly the state of the walker at the first and the
second time step:
\begin{equation}
\begin{aligned}
|\psi(1)\ra &=\frac{1}{\sqrt{|a|^2+|b|^2}}\Big(\big[ a \cos\theta
|\uarr\ra+b\sin\theta|\uarr\ra\big]\otimes|n_0+1\ra  \\
&\hskip90pt +\big[-a \sin\theta
|\darr\ra+b\cos \theta|\darr\ra\big]\otimes|n_0-1\ra\Big), \\
\label{eq:qua-qrw-psi12} \\
|\psi(2)\ra&=\frac{1}{\sqrt{|a|^2+|b|^2}}\Big(\big[a \cos^2\theta
|\uarr\ra+b\cos \theta \sin \theta|\uarr\ra\big]\otimes |n_0+2\ra  \\[-1ex]
&\hskip90pt +\big[-b\sin^2\theta |\darr\ra-a\cos
\theta \sin \theta |\darr\ra\big]\otimes|n_0\ra \\
&\hskip90pt +\big[-a \sin^2\theta
|\uarr\ra+b\cos \theta \sin\theta|\uarr\ra\big]\otimes|n_0\ra  \\
&\hskip90pt +\big[-a \cos \theta \sin \theta |\darr\ra+b\cos^2\theta
|\darr\ra\big]\otimes|n_0-2\ra\Big). 
\end{aligned}
\end{equation}
It is evident then that the state at time $t$ has the general structure
\begin{align}
|\psi(t)\ra=|\uarr\ra\otimes\sum_{n}\Psi_{\rm u}(n,t)|n\ra+|\darr\ra\otimes\sum_{n}\Psi_{\rm d}(n,t)|n\ra,
\label{eq:qua-psi-n}
\end{align}
where the sum extends over all the lattice sites. Here, $\Psi_{\rm
u}(n,t)$ and $\Psi_{\rm d}(n,t)$ are respectively the
probability amplitude to find the random walker on site $n$ at time $t$
with spin state up (u) and down (d). Normalization of $|\psi(t)\ra$
implies $\sum_n P_n(t)=1~\forall~t$, with
\begin{align}
P_n(t) \equiv | \Psi_{\rm u}(n,t)|^2+|\Psi_{\rm d}(n,t)|^2 \label{eq:qrw-site-occu-prob}
\end{align}
being the site occupation probability
for the random walker to be on site $n$ at time $t$.

For $\theta=0, \pi$, an inspection of Eq.~\eqref{eq:qua-qrw-psi12}
suggests that at any time step $t$, only the extreme possible sites to the left
and to the right are going to be occupied. For $\theta=\pi/2, 3\pi/2$,
Eq.~\eqref{eq:qua-qrw-psi12} implies that the random walker is going to
occupy sites $(n_0\pm 1)$ at odd time steps and only the initial site $n_0$ at
even time steps. These scenarios are a result of the operator $C$
effectively implementing no mixing of the spin states $|\uarr \ra$ and
$|\darr \ra$ for $\theta=0, \pi$, and a complete mixing implementing the
transformations $|\uarr \ra \to - |\darr \ra$ and $|\darr \ra \to |\uarr
\ra$ for $\theta = \pi/2$, and $|\uarr \ra \to |\darr \ra$ and $|\darr
\ra \to - |\uarr
\ra$ for $\theta =3\pi/2$. Only for other values of $\theta$, when there is partial mixing, do we get non-trivial results in which several sites are
occupied at any time instant. In
Table~\ref{table:qrw}, we have listed possibilities for nonzero $P_n(t)$ for different choices of the initial location of the
QRW and for $N$ even and odd.
\begin{table*}
\relscale{0.6}
\centering
\begin{tabular}{|p{2.7cm} | p{2.7cm} | p{2.7cm} | p{1.75cm} | p{2.7cm} | p{1.75cm}|}
\hline
\multicolumn{2}{|c|}{$N$ even} & \multicolumn{4}{c|}{$N$ odd}\\
\hline  \hline
\multicolumn{1}{|c|}{Initial site $n_0$ even} &
        \multicolumn{1}{c|}{Initial site $n_0$ odd} &
        \multicolumn{2}{c|}{Initial site $n_0$ even} &
        \multicolumn{2}{c|}{Initial site $n_0$ odd} \\
\hline
\multicolumn{1}{|c|}{For all $t$} & \multicolumn{1}{c|}{For all $t$} & \multicolumn{1}{c|}{$t<N$} & \multicolumn{1}{c|}{$t > N$} & \multicolumn{1}{c|}{$t<N$} & \multicolumn{1}{c|}{$t > N$} \\
\cline{1-6}
\shortstack[l]{\\ For even $t$: nonzero \\ $P_n(t)$ only for even $n$. \\ For odd $t$: nonzero \\ $P_n(t)$ only for odd $n$.} &
\shortstack[l]{\\ For even $t$: nonzero \\$P_n(t)$ only for odd $n$. \\ For odd $t$:  nonzero \\ $P_n(t)$ only for even $n$.} &
\shortstack[l]{\\ For even $t$: nonzero \\ $P_n(t)$ only for even $n$. \\ For odd $t$: nonzero \\ $P_n(t)$ only for odd $n$.} & 
\shortstack[l]{\\ Non-zero $P_n(t)$ \\ possible for all \\ $n$ and $t$.  {} \\ {} \\ {} } & 
\shortstack[l]{\\ For even $t$: nonzero \\ $P_n(t)$ only for odd $n$. \\ For odd $t$: nonzero \\ $P_n(t)$ only for even $n$.} & 
\shortstack[l]{\\ Non-zero $P_n(t)$ \\ possible for all \\ $n$ and $t$.  {} \\ {} \\ {} } \\
\hline
\end{tabular}
\caption{Site occupation probability $P_n(t)$ in the case of the QRW on a
one-dimensional periodic lattice with $N$ sites. Here, the initial state
is given  by Eq.~\eqref{eq:qua-psi0}.}
\label{table:qrw}
\end{table*}

We now present analytical results on the site occupation probability $P_n(t)$ for the
initial state~\eqref{eq:qua-psi0}. To proceed, we note that acting on
the state~\eqref{eq:qua-psi-n} by $C$ followed by $U$ gives
linear equations expressing $\Psi_{\rm u,d}(n,t+1)$ in terms of
$\Psi_{\rm u,d}(n-1,t)$ and $\Psi_{\rm u,d}(n+1,t)$, which may be put in the form of a matrix
equation by introducing the two-component vector~\cite{qua-Vishwanath:2000}
\begin{align}
\ket{\Psi(n,t)} \equiv [\Psi_{\rm u}(n,t)~~\Psi_{\rm d}(n,t)]^{\mathsf{T}} \, . \label{eq:qrw-two-comp-vector}
\end{align}
Here, $\mathsf{T}$ denotes
transpose operation. Then, the matrix equation reads
\begin{align}
\ket{\Psi(n,t+1)}
&= \begin{bmatrix} \cos \theta  &\sin \theta \\ 0 & 0 \end{bmatrix}
\ket{\Psi(n-1,t)} + \begin{bmatrix} 0 & 0 \\ -\sin \theta & \cos \theta \end{bmatrix} \ket{\Psi(n+1,t)}.
\label{eq:qua-Psi-eqn}
\end{align}

To solve Eq.~\eqref{eq:qua-Psi-eqn}, we employ the method of discrete Fourier
transform~\cite{qua-Vishwanath:2000}, to write $\ket{\tilde\Psi(k,t)} =
\sum_{n}
\ket{\Psi(n,t)}\exp(\ii 2\pi kn/N)$ and $\ket{\Psi(n,t)} = (1/N) \sum_{k}
\ket{\tilde\Psi(k,t)} \exp(-\ii 2\pi kn/N)$,
with $n,k \in [-(N-1)/2,(N-1)/2]$ if $N$ is odd, and $n,k \in [-N/2,N/2-1]$ if $N$ is even.
Equation~\eqref{eq:qua-Psi-eqn} then gives 
\begin{align}
\ket{\tilde\Psi(k,t)}= (M_k)^t ~\ket{\tilde\Psi(k,0)},
\label{eq:qua-iteration} 
\end{align}
where the matrix $M_k$ is given by 
\begin{align}
M_k &\equiv \begin{bmatrix} \ee^{\ii 2\pi k/N}\cos\theta &
\ee^{\ii 2\pi k/N}\sin\theta \\ - \ee^{-\ii 2\pi k/N}\sin\theta &
\ee^{-\ii 2\pi k/N}\cos\theta \end{bmatrix}  = \lambda^{(1)}_k \ket{\phi^{(1)}_{k}} \bra{\phi^{(1)}_{k}} +
\lambda^{(2)}_k \ket{\phi^{(2)}_{k}} \bra{\phi^{(2)}_{k}},
\label{eq:qua-Mk-matrix}
\end{align}
in terms of its eigenvalues and orthonormal eigenvectors:
\begin{equation}
\begin{aligned}
& \lambda^{(1)}_{k} = \ee^{\ii\omega_k}  \, , &\ket{\phi^{(1)}_{k}} = \frac{1}{\sqrt{1+h^2_{+}(k)}}
\begin{bmatrix} -\ii \ee^{\ii 2\pi k/N} h_{+}(k) \\ 1
\end{bmatrix}, \\[1ex] 
& \lambda^{(2)}_{k} = \ee^{-\ii\omega_k}  \, , &\ket{\phi^{(2)}_{k}} = \frac{1}{\sqrt{1+h^2_{-}(k)}}
\begin{bmatrix} -\ii \ee^{\ii 2\pi k/N} h_{-}(k) \\ 1
\end{bmatrix},  
\end{aligned}
\end{equation}
with
$\cos{\omega_k} \equiv  \cos{\left(2\pi k/N\right)}\cos\theta$, and $h_{\pm}(k)
\equiv  \cot\theta \sin{\left(2\pi k/N\right)} \pm
\csc\theta \sin\omega_k$. Note that $h_{+}(-k)=-h_{-}(k)$, and $h_{+}(k)h_{-}(k)=-1$. The initial
condition corresponding to Eq.~\eqref{eq:qua-psi0} is
$\ket{\tilde{\Psi}(k,0)} = \left({\exp{(\ii 2\pi k n_0/
N)}}/\sqrt{|a|^2+|b|^2}\right) [a~~b]^{\mathsf{T}}~\forall~k$.
Using these results to obtain $\ket{\tilde\Psi(k,t)}=(\lambda^{(1)}_k)^t  ~\bk{\phi^{(1)}_{k}}{\tilde\Psi(k,0)}
 ~\ket{\phi^{(1)}_{k}} + (\lambda^{(2)}_k)^t~
 \bk{\phi^{(2)}_{k}}{\tilde\Psi(k,0)} ~\ket{\phi^{(2)}_{k}}$, and then performing inverse Fourier transformation, we finally obtain for odd $N$ the result
\begin{align}
\Psi_{\rm u}(n,t)=\hspace{-0.5cm}\sum^{(N-1)/2}_{k=-(N-1)/2}
~{\cal A}_{\rm u}(k,n,t),~\Psi_{\rm
d}(n,t)=\hspace{-0.5cm}\sum^{(N-1)/2}_{k=-(N-1)/2}{\cal A}_{\rm d}(k,n,t),
\label{eq:qua-psiud-final}
\end{align}
while for even $N$, we get
\begin{equation}
\begin{aligned}
 \Psi_{\rm u}(n,t) &= (-1)^{n-n_0 +t}\frac{\big\{
 a\cos(\theta t)+b\sin(\theta t)  \big\}}{N\sqrt{|a|^2+|b|^2}}  +\sum^{N/2-1}_{k=-N/2+1}{\cal
 A}_{\rm u}(k,n,t),  \\[1ex]
\label{eq:qua-psiud-even-final}
 \Psi_{\rm d}(n,t) &= (-1)^{n -n_0 +t}\frac{\big\{
 -a\sin(\theta t) + b\cos(\theta t) \big\}}{N\sqrt{|a|^2+|b|^2}}
 + \sum^{N/2-1}_{k=-N/2+1}{\cal
 A}_{\rm d}(k,n,t).
\end{aligned}
\end{equation}
Here, the quantities ${\cal A}_{\rm u}(k,n,t)$ and ${\cal A}_{\rm
d}(k,n,t)$ are given by 
\begin{equation}
\begin{aligned}
        {\cal A}_{\rm u}(k,n,t) &=  {\cal N}(a,b,k) \big[a\cos \{ 2\pi k(n -n_0)/N+\omega_k t
        \} \\
        &\hskip70pt +bh_{+}(k) \sin\left\{2\pi k(n-n_0 -1)/N+\omega_k t\right\} \big], \\[1ex]
        {\cal A}_{\rm
        d}(k,n,t) &= {\cal N}(a,b,k) \big[- ah_{+}(k) \sin\left\{-2\pi k(n-n_0
        +1)/N+\omega_k t\right\} \\
        &\hskip70pt+b\cos\{-2\pi k (n-n_0) /N+\omega_k
t\}\big], 
\end{aligned}
\end{equation}
with ${\cal N}(a,b,k)=
2/\left[N(1+h^2_+(k))\sqrt{|a|^2+|b|^2}\right]$. 

Let us remark that implementing the transformation $\theta \to \theta+\pi$ in
the expression for $C$ is tantamount to multiplying $C$ and consequently
the matrix $M_k$ by the factor $-1$. As a result, the eigenvalues of
$M_k$ get
both multiplied by the factor $-1$, although the corresponding
eigenvectors remain the same. All of these
would however leave $|\Psi_{\rm
u}(n,t)|^2$ and $|\Psi_{\rm d}(n,t)|^2$ and consequently the site
occupation probability $P_n(t)$ unchanged, thereby making us
conclude that the QRW is invariant with respect to
the transformation $\theta \to \theta+\pi$. Hence, we will in the rest
of the paper restrict the values of $\theta$ to the range $[0,\pi]$.

\begin{figure}[!htbp]
\centering
\includegraphics[scale=0.85]{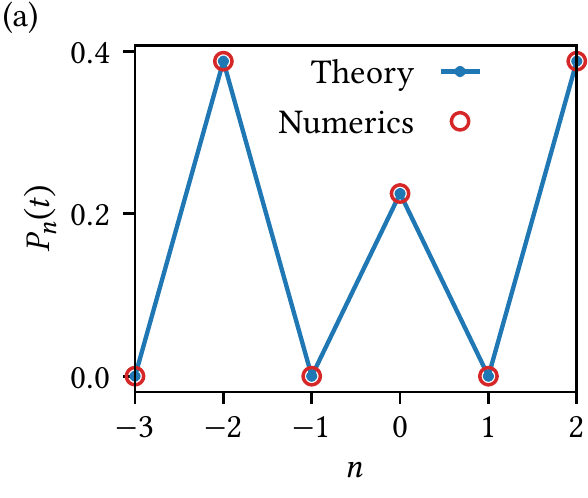}\hskip20pt
\includegraphics[scale=0.85]{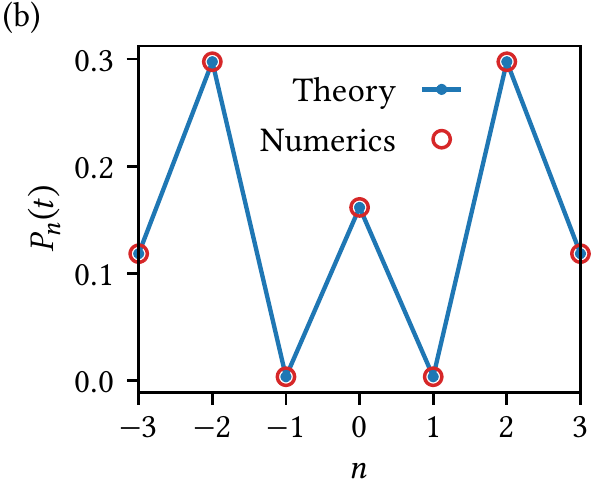}\\[0.1ex]
\includegraphics[scale=0.85]{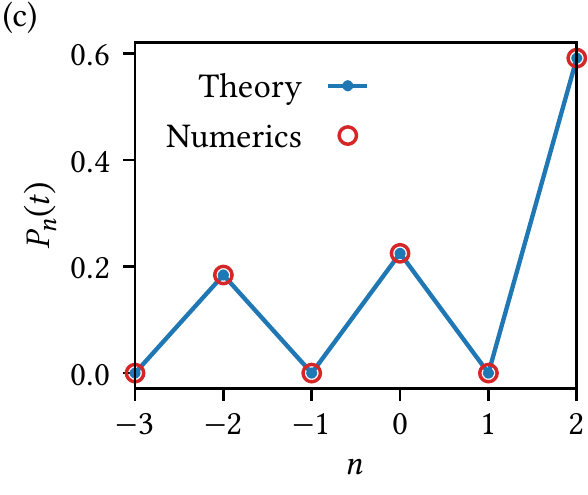}\hskip20pt
\includegraphics[scale=0.85]{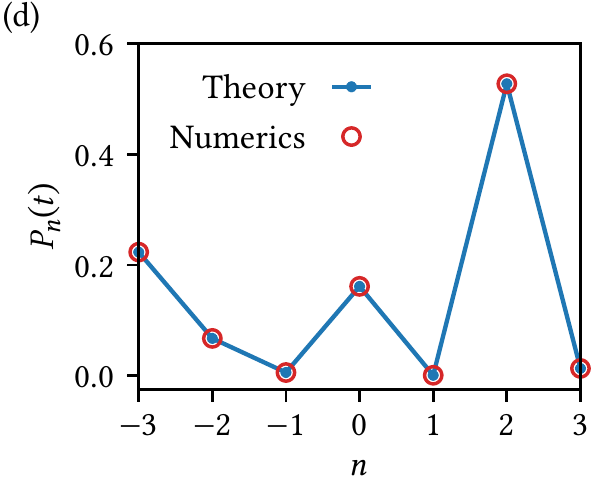}
        \caption{Site occupation probability $P_n(t)$ for the QRW on
a one-dimensional periodic lattice of $N$ sites and at time $t=20$, while starting
from the state~\eqref{eq:qua-psi0} with $n_0=0$. The values of the
        parameters $a$ and $b$ defining the initial state are
        $a=1,~b=\ii$ for panels (a) and (b), and $a=b=1$ for panels (c)
        and (d). The value of $N$ is $N=6$ for panels (a) and (c), and
        $N=7$ for panels (b) and (d). The angle $\theta$ has a
        value in radian that corresponds to $80$ degrees. Numerics
correspond to bare evolution of the unitary dynamics of the model, while
analytical results are given by Eqs.~\eqref{eq:qrw-site-occu-prob},~\eqref{eq:qua-psiud-final}
and~\eqref{eq:qua-psiud-even-final}. The lines are a guide to the eye.}
\label{fig:qua-qrw-site-probability}
\end{figure}

In Fig.~\ref{fig:qua-qrw-site-probability}, we show a comparison between
numerical results and theory for the site occupation probability $P_n(t)$, demonstrating a perfect match.
In the figure, numerics
correspond to bare evolution of the unitary dynamics of the model, while
analytical results are given by Eqs.~\eqref{eq:qrw-site-occu-prob},~\eqref{eq:qua-psiud-final}
and~\eqref{eq:qua-psiud-even-final}. 

\subsection{Random projective measurements and survival probability}
\label{sec:QRW3}

\subsubsection{Numerical results}
\label{sec:QRW3a}

We now present numerical results on the survival
probability $S_m$, and in particular, on the average and the typical
survival probability, and their dependence on the number of measurements $m$.  To proceed, we note in the context of the discussion on measurement schemes in Section \ref{sec:Introduction} that the time intervals $\tau_\alpha$ in the present case are positive integers with underlying discrete distribution $p_\tau$~\cite{ptau}.  Moreover,  since the QRW evolves in discrete times, the unitary operator $U_\alpha$ is given by $U_\alpha \equiv [U (C \otimes \mathbb{I})]^{\tau_\alpha}$.  Thus, a single-step evolution of the QRW  
involves acting by the operator $C \otimes \mathbb{I}$ followed by
the operator $U$, see Eqs.~\eqref{eq:qua-U-definition}
and~\eqref{eq:qua-Hadamard-definition}.  Since the QRW evolves in
discrete times, the phrases ``time" and ``time step" would have the same
meaning in the context of the QRW and would be used interchangeably in
the following.

As a representative case for reporting our results, we take the QRW lattice size $N$ to be
even with $n_0=0$. It then follows from Table~\ref{table:qrw} that the
site occupation probability for our choice of the initial state is
nonzero on site $0$ only at even time steps. Consequently, we must choose
the i.i.d.~random variables $\tau_\alpha$ as even numbers, as otherwise any
projective measurement on the instantaneous state would yield null
result. For our purpose, we make for the $\tau_\alpha$ distribution $p_\tau$
two representative choices, namely, that of a discrete
exponential distribution
\begin{equation}
\begin{aligned}
\label{eq:qua-app-exp}
& p_{\tau} = r(1-r)^{\tau/2-1};~\quad \tau=2,4,6,\ldots,~\quad 0 < r < 1 \, ; \\
& \langle \tau \rangle = \frac{2}{r} \, , \quad {\mathrm{Var}}[\tau] = \frac{4(1-r)}{r^2} \, ,
\end{aligned}
\end{equation}
and that of a power-law distribution 
\begin{equation}
\begin{aligned}
\label{eq:qua-app-pow}
& p_{\tau}= \frac{2^s}{\zeta(s) \tau^s};~\quad \tau=2,4,6,\ldots,~\quad s > 1\, ;  \\
& \langle \tau \rangle =  \frac{ 2 \,\zeta(s-1)}{\zeta(s)} \, , ~ s>2 \, ;\quad {\mathrm{Var}}[\tau] = \frac{4\big[\zeta(s)\zeta(s-2)-\zeta^2(s-1)\big]}{\zeta^2(s)} \, ,~ s>3 \, ;
\end{aligned}
\end{equation}
where $\zeta(s)\equiv \sum_{n=1}^\infty 1/n^s ; ~s>1$ is the Riemann zeta function. Here,  $\langle \tau \rangle$ denotes the mean of the distribution $p_\tau$, while $\mathrm{Var}[\tau]$ denotes its variance.  All the aforementioned distributions satisfy the normalization $\sum_{\tau=2,4,6,\ldots}^{\infty} p_{\tau} = 1$.
The exponential distribution~\eqref{eq:qua-app-exp} has all its moments,
and in particular, the mean finite for all values
of the parameter $r$. By contrast, the power-law distribution~\eqref{eq:qua-app-pow} has a finite mean only
for $s>2$ and a finite variance only for $s>3$. Since any
reasonable experimental set-up would allow measurements to be performed
at random time intervals that have a finite average, as argued above, we will in this work
consider only values of $s$ larger than $2$. 

\begin{figure*}[!ht]
\centering
\includegraphics[scale=0.78]{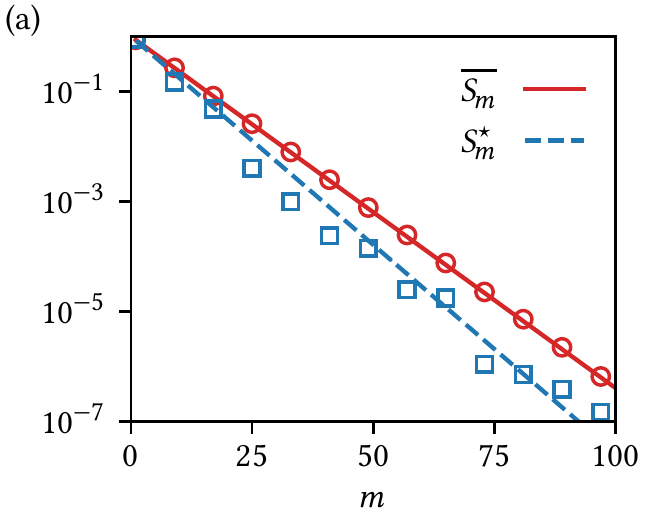} 
\includegraphics[scale=0.78]{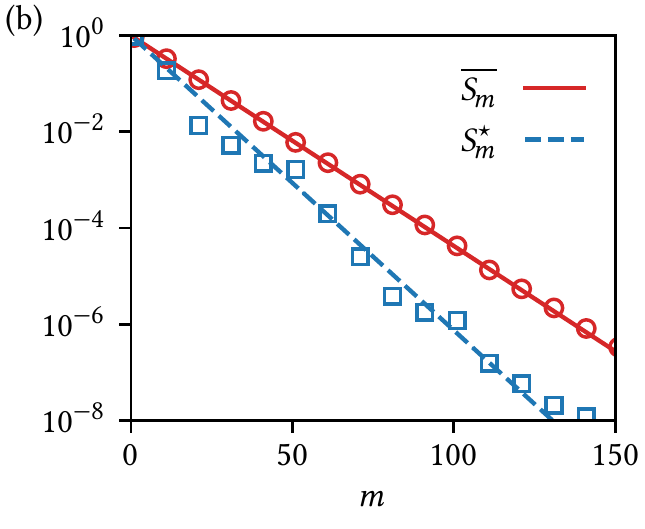} 
\includegraphics[scale=0.78]{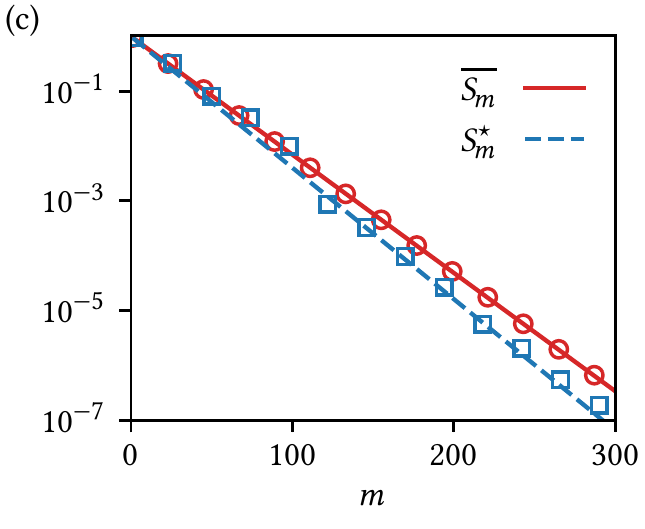}
\caption{Average and typical survival probability for the QRW subject to instantaneous projective measurements at random times
(\textbf{Scheme 1}). The plots correspond to the initial state~\eqref{eq:qua-psi0}
with $n_0=0,~a=1,~b={\rm i}$ that is subject to repeated projective
measurements at random times to the initial state and subsequent evolution with the
projected component of the instantaneous state. Here, the time intervals $\tau_\alpha$ between two consecutive
measurements are i.i.d.~random variables sampled from the exponential
distribution~\eqref{eq:qua-app-exp} with $r=0.5$ (panel (a)) and
from the power-law distribution~\eqref{eq:qua-app-pow} with
$s=2.5$ (panel (b)) and $s=3.5$ (panel (c)). The angle $\theta$ characterizing the
        QRW evolution operator $C$ has the
        value in radian corresponding to $80$ degrees. The system size
        is $N=500$. In the plots, the points are based on results
        obtained from numerical implementation of the dynamics; while
        the average survival probability $\overline{S_m}$ involves
        averaging over $3000$ realizations of the set $\{\tau_\alpha\}_{1\le
        \alpha \le m}$, the typical survival probability $S_m^\star$
        corresponds to results obtained in a typical realization of the
        $\tau_\alpha$'s. The lines in the plots correspond to analytical
        results given by
        Eqs.~\eqref{eq:qua-Sm-avg-qrw},~\eqref{eq:qua-Sm-typical-qrw},
        and~\eqref{eq:qua-qrw-qtau}.}  
\label{fig:qua-qrw-scheme1}
\end{figure*}

In Fig.~\ref{fig:qua-qrw-scheme1}, we show in case of the evolution
under \textbf{Scheme 1} our numerical results on the average survival
probability $\overline{S_m}$ when averaged over typically hundreds of
realizations $\{\tau_\alpha\}_{1\le \alpha \le m}$ and the survival probability $S_m^\star$
obtained in a typical realization of the $\tau_\alpha$'s, both plotted as a
function the $m$, the number of measurements.  In the figure, panel (a)
corresponds to $\tau_\alpha$'s distributed according to the exponential
distribution~\eqref{eq:qua-app-exp} with $r=0.5$, while panels (b) and (c) are for
the power-law distribution~\eqref{eq:qua-app-pow} with $s$ having values
$2.5$ and $3.5$, respectively. The panels suggest for both
the quantities $\overline{S_m}$ and $S_m^\star$ an
exponential decay with $m$ for large $m$.

\begin{figure}[!htbp]
\centering
\includegraphics[scale=0.78]{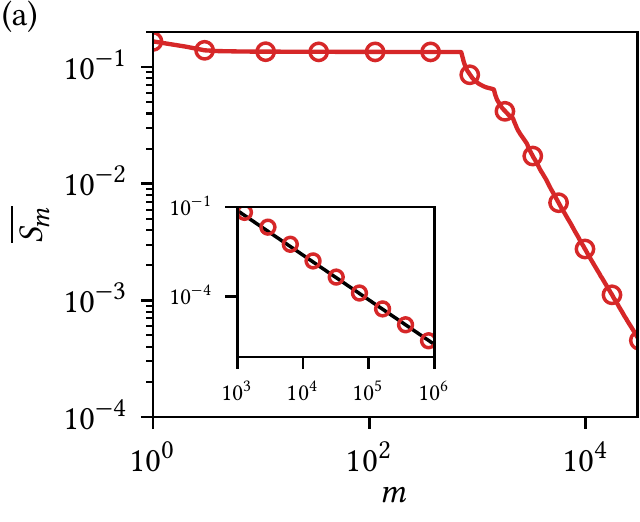} \hskip1pt
\includegraphics[scale=0.78]{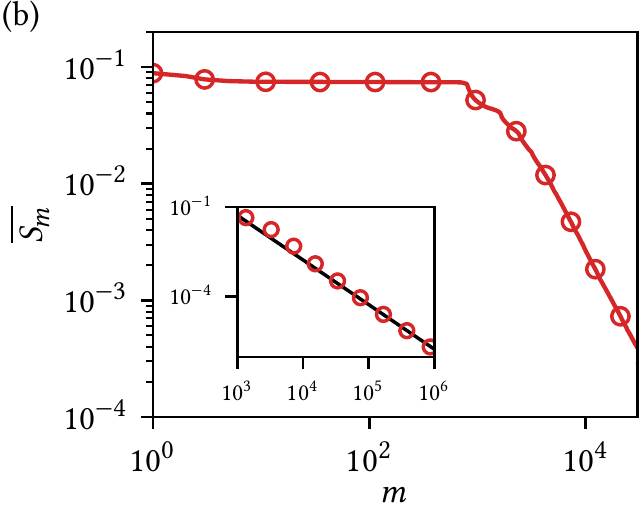} \hskip1pt
\includegraphics[scale=0.78]{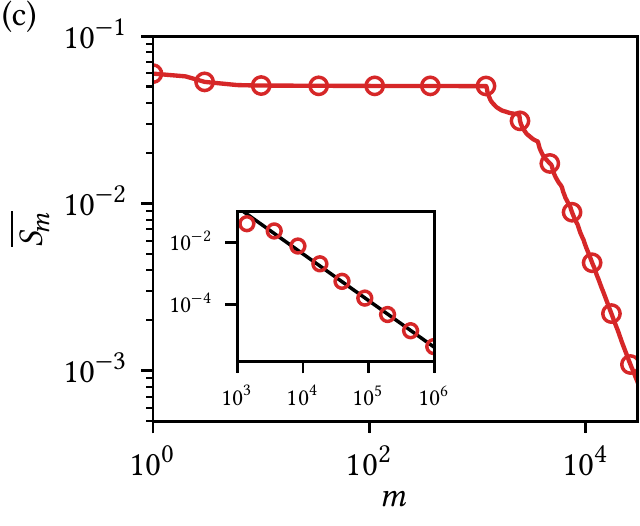} \\[2ex]
\includegraphics[scale=0.78]{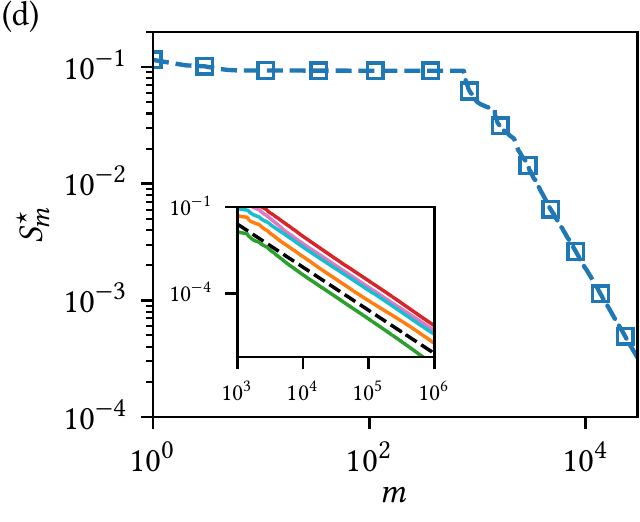} \hskip1pt
\includegraphics[scale=0.78]{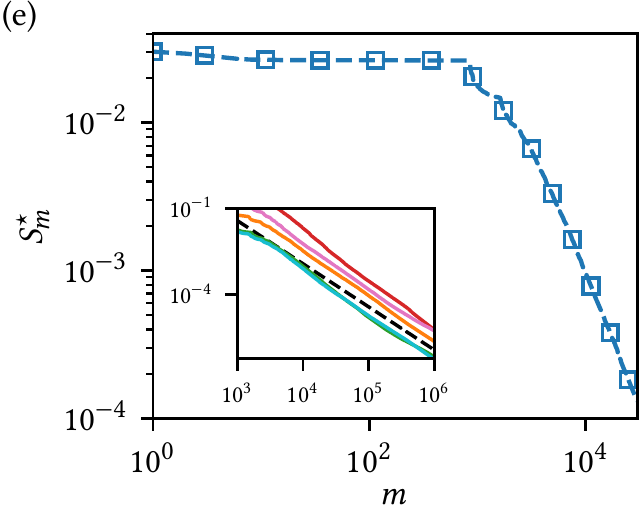} \hskip1pt
\includegraphics[scale=0.78]{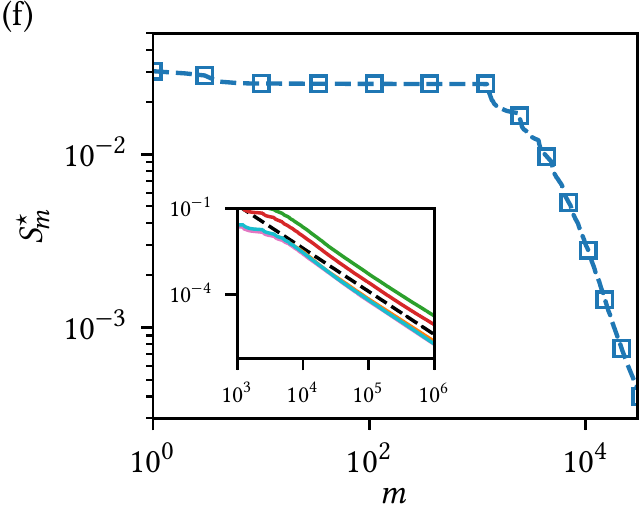}
        \caption{Average survival probability (panels (a) -- (c)) and
        typical survival probability (panels (d) -- (f)) for the QRW subject to instantaneous projective measurements at random times
(\textbf{Scheme 2}). The plots correspond to the initial state~\eqref{eq:qua-psi0}
with $n_0=0,~a=1,~b={\rm i}$ that is subject to repeated projective
measurements at random times to the initial state and subsequent
        evolution with the leftover component of the instantaneous
        state after the measurement. Here, the time intervals $\tau_\alpha$ between two consecutive
measurements are i.i.d.~random variables sampled from the exponential
        distribution~\eqref{eq:qua-app-exp} with $r=0.5$ (panels (a) and
        (d)), and
from the power-law distribution~\eqref{eq:qua-app-pow} with
        $s=2.5$ (panels (b) and (e)) and $s=3.5$ (panels (c) and (f)). The angle $\theta$ characterizing the
        QRW evolution operator $C$ has the
        value in radian corresponding to $80$ degrees. The system size
        is $N=500$. In the main plots, the points are based on results
        obtained from numerical implementation of the dynamics; while
        the average survival probability $\overline{S_m}$ involves
        averaging over $10$ realizations (panel (a)), over $20$ realizations (panel (b)), and over $10$ realizations (panel (c)) of the set $\{\tau_\alpha\}_{1\le
        \alpha \le m}$, the typical survival probability $S_m^\star$
        corresponds to results obtained in a typical realization of the
        $\tau_\alpha$'s. The lines in the main plots correspond to those
        obtained from the semi-analytical approach described in the
        text, see Section~\ref{sec:QRW3b}. In the insets in the upper
        row, the points correspond to numerically-evaluated average
        survival probability, while the line represents an $m^{-3/2}$
        behavior. In the insets in the lower
        row, the continuous lines correspond to numerically-evaluated
        survival probability for five typical realizations of
        $\{\tau_\alpha\}_{1\le \alpha \le m}$, while the dashed line
        represents an $m^{-3/2}$ behavior. We conclude from the insets
        in both the upper and the lower row that the average as well as the typical
        survival probability behaves at large $m$ as $m^{-3/2}$.}
\label{fig:qua-qrw-scheme2}
\end{figure}

\begin{figure}[!htbp]
\centering
\includegraphics[scale=1]{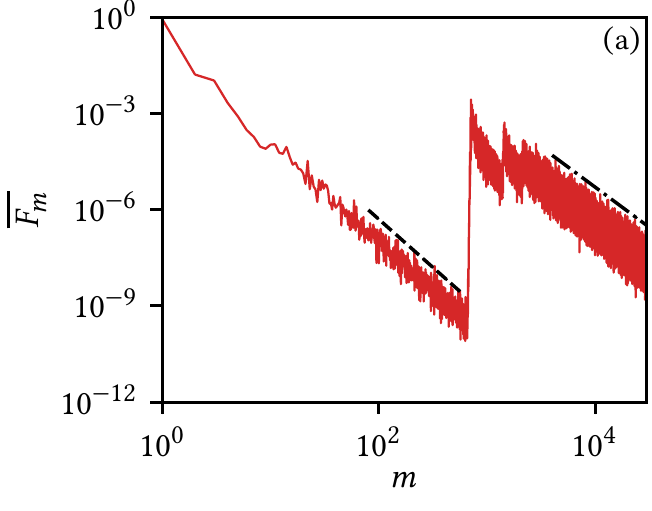} \hskip5pt 
\includegraphics[scale=1]{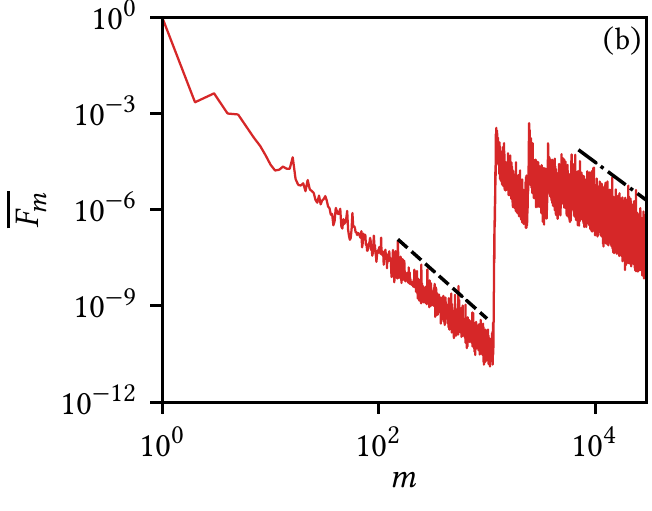}   
\caption{Average first-detection probability for the QRW model (\textbf{Scheme 2}).  For panels (a) and (b), parameter values and number of averaging realizations are the same as in Figs.~\ref{fig:qua-qrw-scheme2}(a) and~\ref{fig:qua-qrw-scheme2}(c), respectively.  As shown in the plots, one may observe two distinct behaviours $\sim m^{-3}$ (dashed line) and $\sim m^{-5/2}$ (dash-dotted line).}
\label{fig:qrw-f-vs-m-exp-pow}
\end{figure}

\begin{figure}[!htbp]
\centering
\includegraphics[scale=1]{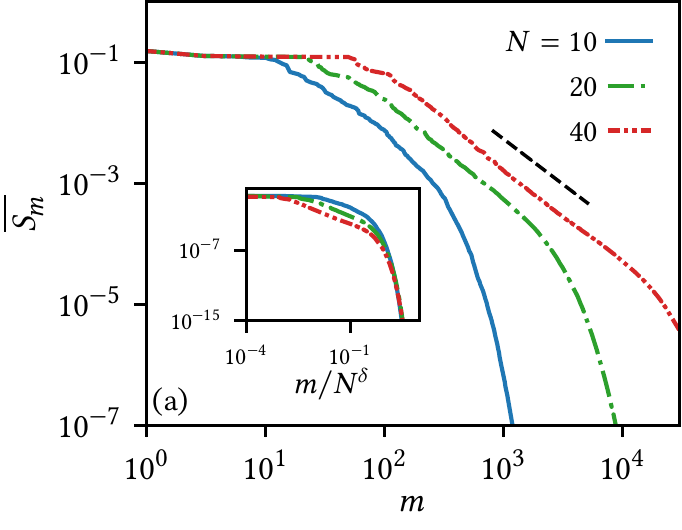} \hskip5pt 
\includegraphics[scale=1]{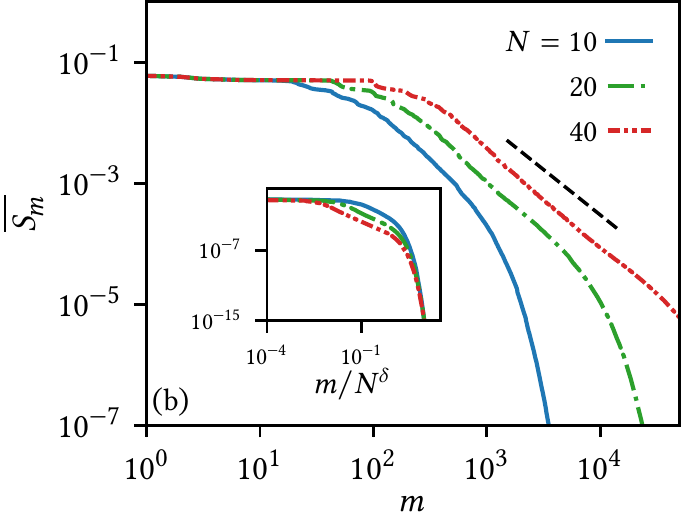}   
\caption{Average survival probability for the QRW model (\textbf{Scheme 2}).  For panels (a) and (b), parameter values other than $N$ and the number of averaging realizations are the same as in Figs.~\ref{fig:qua-qrw-scheme2}(a) and~\ref{fig:qua-qrw-scheme2}(c), respectively.  The main plots show a crossover from a $m^{-3/2}$-behaviour (dashed line) to an exponential tail over the characteristic value $m_2^\star(N)$ of $m$; the collapse of the data shown in the insets suggests the scaling~\eqref{eq:m2-scaling-QRW}.}
\label{fig:qrw-N-dependence}
\end{figure}

\begin{figure}[!htbp]
\centering
\includegraphics[scale=1.]{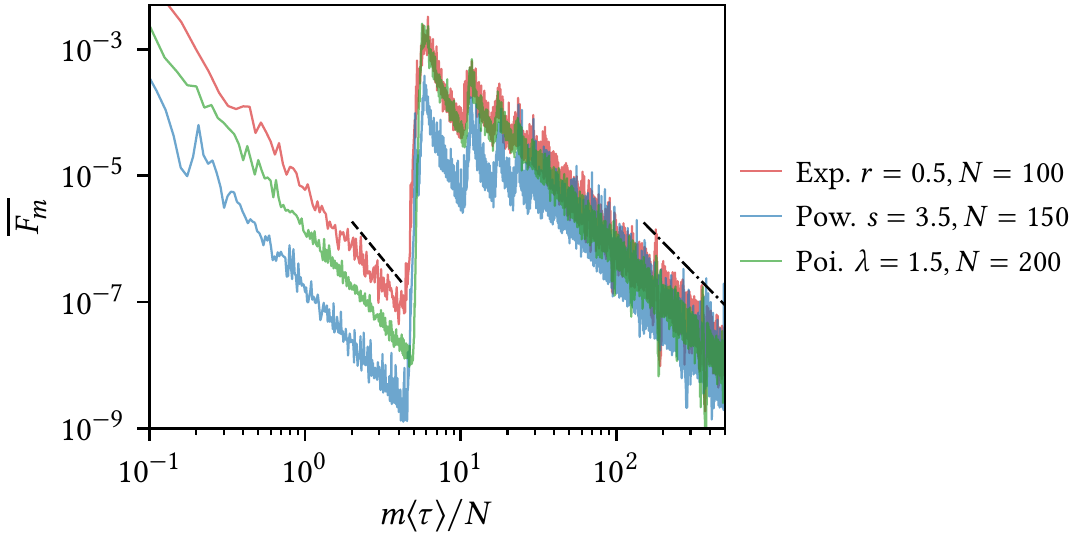} 
\caption{Average first-detection probability for the QRW model (\textbf{Scheme 2}) for the exponential distribution~(\ref{eq:qua-app-exp}), the power-law distribution~(\ref{eq:qua-app-pow}), and the Poisson distribution~(\ref{eq:qua-Poisson}).  Here, the parameters $n_0$,  $a$,  $b$, and $\theta$ have the same values as in Fig.~\ref{fig:qua-qrw-scheme2}.  The number of averaging realizations in all cases is $50$.  The parameter $r$ for the exponential distribution and the parameter $s$ for the power-law distribution have the same values as in Fig.~\ref{fig:qua-qrw-scheme2}, panels (a) and (c), respectively.  
For the data plotted for the Poisson distribution~(\ref{eq:qua-Poisson}), we have $\lambda=1.5$.
The plot suggests the scaling depicted in Eq.~(\ref{eq:m1-scaling-QRW}). }
\label{fig:qrw-fm}
\end{figure}

We now present our results for {\bf Scheme 2}. In order to contrast with those presented in Fig.~\ref{fig:qua-qrw-scheme1}\, for {\bf Scheme 1}, we use the same initial state.  Figure~\ref{fig:qua-qrw-scheme2}\, 
shows our numerical results on the average survival probability
$\overline{S_m}$ and the typical survival probability $S_m^\star$ for the exponential
distribution~\eqref{eq:qua-app-exp} with $r=0.5$ (panels (a) and (d)),
and for the power-law distribution~\eqref{eq:qua-app-pow} with $s=2.5$ (panels (b) and (e)) and $s=3.5$ (panels (c) and (f)).  Based on our obtained results, we first summarize the behavior of $\overline{S_m}$.
\begin{enumerate}
\item There are two characteristic $N$-dependent $m$ values, namely, $m_1^\star(N)$ and $m_2^\star(N)$, that characterize the behaviour of $\overline{S_m}$.  While $m_1^\star(N)$ scales linearly with $N$, the other scale $m_2^\star(N)$ grows superlinearly with $N$.
\item For $m < m_1^\star(N)$ and large, one has $\overline{S_m} \sim m^{-2}$, implying  thereby the behavior of the average first-detection probability as $\overline{F_m} \sim m^{-3}$,  see Eq.~(\ref{eq:Fm-definition}) for the definition of $F_m$ in terms of $S_m$.  
\item For $m$ satisfying $m_1^\star(N) \ll m < m_2^\star(N)$, one has $\overline{S_m} \sim m^{-3/2}$, implying  thereby the behavior of the average first-detection probability as $\overline{F_m} \sim m^{-5/2}$.  
\item For $m \gg m_2^\star(N)$, one has an exponential decay of $\overline{S_m}$ with $m$.
\item While the $m^{-3/2}$-behavior of $\overline{S_m}$ may already be observed in Fig.~\ref{fig:qua-qrw-scheme2},  in the insets of panels (a) -- (c),  in order to observe the $m^{-2}$-behaviour, it proves convenient to look at the plot of the corresponding $\overline{F_m}$, see Fig.~\ref{fig:qrw-f-vs-m-exp-pow}, in which the two aforementioned distinct behaviours of $m^{-3}$ and $m^{-5/2}$ may be clearly observed.  The exponential decay of $\overline{S_m}$ with $m$ for $m \gg m_2^\star(N)$ may be seen from Fig.~\ref{fig:qrw-N-dependence}.  Note that since $m_2^\star(N)$ grows superlinearly with $N$, in order to observe this exponential decay for the value of $N$ used in Fig.~\ref{fig:qua-qrw-scheme2}, one has to obtain results for very large values of $m$,  requiring computation for a prohibitively-long time.  Hence, we use in Fig.~\ref{fig:qrw-N-dependence}\, a smaller $N$ than the one used  in Fig.~\ref{fig:qua-qrw-scheme2}.
\item That the characteristic scale $m_1^\star(N)$ scales linearly with $N$ may be deduced from Fig.~\ref{fig:qrw-fm}, which indeed implies that
\begin{align}
m_1^\star(N) \sim \frac{N}{\langle \tau \rangle},
\label{eq:m1-scaling-QRW}
\end{align}
where $\langle \tau\rangle $ is the average of the distribution $p_\tau$. 
\item The superlinear scaling of $m_2^\star(N)$ with $N$ may be deduced from the insets of Fig.~\ref{fig:qrw-N-dependence}.  We find: 
\begin{align}
m_2^\star(N) \sim N^\delta;~\quad\delta=3.0.
\label{eq:m2-scaling-QRW}
\end{align}
\end{enumerate}
In the above backdrop, we may mention that if one studies an infinite system (i.e.,  the limit $N \to \infty$),  the scale $m_1^\star(N)$ diverges,  and one would observe  the large-$m$ behavior $\overline{S_m} \sim m^{-2}$ and $\overline{F_m} \sim m^{-3}$.

\begin{figure}[!htbp]
\centering
\includegraphics[scale=1.1]{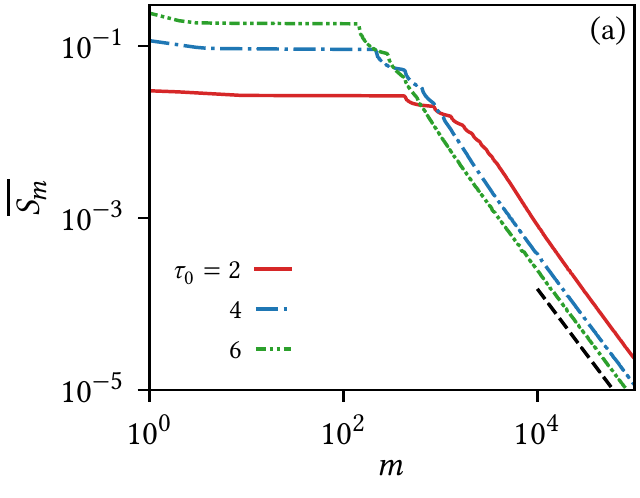} \hskip5pt 
\includegraphics[scale=1.1]{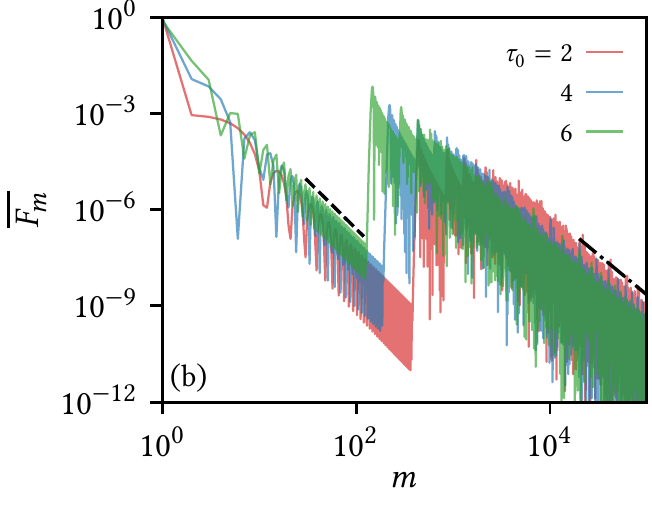}   
\caption{Average survival and first-detection probability for the QRW model (\textbf{Scheme 2}).  Here, the parameters $n_0$,  $a$,  $b$,  and $\theta$ have the same values as in Fig.~\ref{fig:qua-qrw-scheme2}.  Here, the time intervals $\tau_\alpha$ between two consecutive
measurements are i.i.d.~random variables sampled from the
        distribution~(\ref{eq:qua-delta}).  The values of the parameter $\tau_0$ are shown in the figure.  The system size
        is $N=150$.  In panel (a),  as shown in the plot, one may observe a behavior $\sim m^{-3/2}$ (dashed line), while in panel (b),  two distinct behaviours $\sim m^{-3}$ (dashed line) and $\sim m^{-5/2}$ (dash-dotted line) may be seen.  Note that since $\tau$ has only one allowed value, namely, $\tau_0$,  it is redundant to use the overbar over $S_m$ and $F_m$ to denote their average values.}
\label{fig:qrw-samet-sp-fp}
\end{figure}

\begin{figure}[!htbp]
\centering
\includegraphics[scale=1.1]{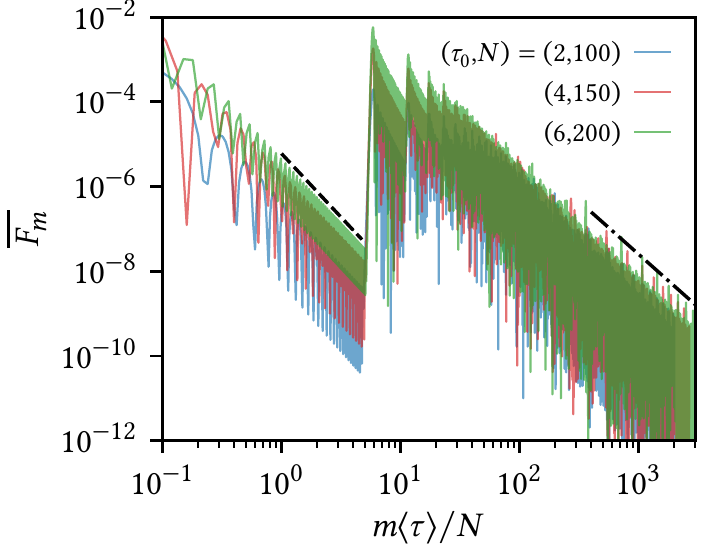}   
\caption{Average first-detection probability for the QRW model (\textbf{Scheme 2}). Parameter values and other details are the same as in Fig.~\ref{fig:qrw-samet-sp-fp}.  The plot suggests that the crossover from the $\sim m^{-3}$-behaviour (dashed line) to the $\sim m^{-5/2}$-behavior (dash-dotted line) over the characteristic value $m_1^\star(N)$ of $m$ that follows the scaling~\eqref{eq:m1-scaling-QRW}.  Note that since $\tau$ has only one allowed value, namely, $\tau_0$,  it is redundant to use the overbar over $F_m$ to denote its average.}
\label{fig:qrw-fp-scaled}
\end{figure}
 
One may wonder about the robustness of all of the aforementioned behaviour with respect to distributions $p_\tau$ other than the exponential distribution~\eqref{eq:qua-app-exp} and the power-law distribution~\eqref{eq:qua-app-pow} that we have considered until now. To this end, we now demonstrate that the same behavior is also observed for two other distributions, namely,  the case corresponding to measurements performed at regular time interval $\tau_0$:
\begin{align}
p_\tau=\delta_{\tau,\tau_0};~\quad \tau=2,4,6,\ldots;~\quad \tau_0>0\, ; \quad \langle \tau \rangle = \tau_0 \, , \quad {\mathrm{Var}}[\tau] = 0 \, ,
\label{eq:qua-delta}
\end{align}
and the Poisson distribution
\begin{align}
p_\tau =    \frac{  \ee^{-\lambda} \, \lambda^{\tau/2-1} }{(\tau/2-1)!} ;~\quad \tau=2,4,6,\ldots;~\quad \lambda > 0\, ; \quad \langle \tau \rangle = 2 (1+\lambda) \, , \quad {\mathrm{Var}}[\tau] = 4 \lambda \, ,
\label{eq:qua-Poisson}
\end{align}
as one varies the parameter $\lambda$ that controls the width of the distribution $p_\tau$, leading to a small (respectively, large) variance for small (respectively, large) $\lambda$.  Results for the case of the delta-function distribution~(\ref{eq:qua-delta}) are shown in Figs.~\ref{fig:qrw-samet-sp-fp} and~\ref{fig:qrw-fp-scaled}.  One may observe from these figures and in appropriate regimes the $m^{-3/2}$ behavior of $\overline{S_m}$, the $m^{-3}$ behavior and the $m^{-5/2}$ behavior of $\overline{F_m}$,  and the scaling of the crossover scale $m_1^\star(N)$ according to Eq.~(\ref{eq:m1-scaling-QRW}).  Since similar results are obtained for the case of the Poisson distribution~\eqref{eq:qua-Poisson},  we do not present all the results here, excepting to show in Fig.~\ref{fig:qrw-fm}\, the scaling of $m_1^\star(N)$ in accordance with Eq.~(\ref{eq:m1-scaling-QRW}).
 
\begin{figure}[!htbp]
\centering
\includegraphics[scale=1]{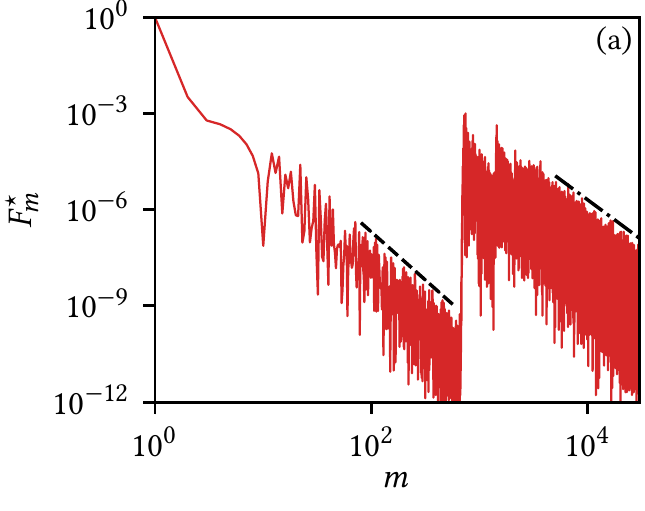} \hskip5pt 
\includegraphics[scale=1]{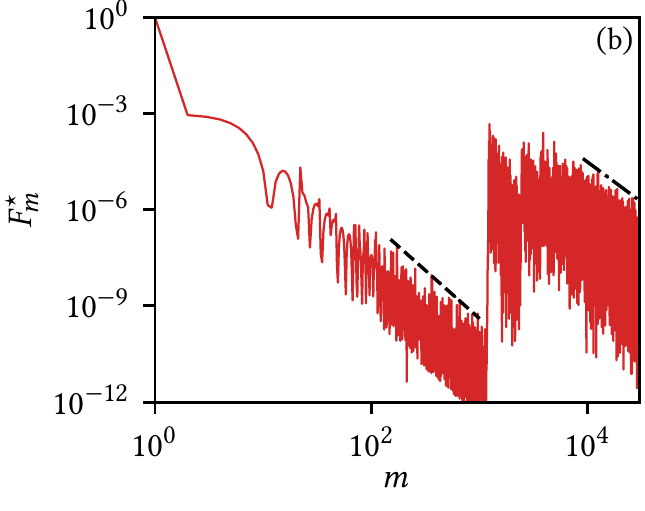}   
\caption{Typical first-detection probability for the QRW model (\textbf{Scheme 2}).  For panels (a) and (b), parameter values are the same as in Figs.~\ref{fig:qua-qrw-scheme2}(a) and~\ref{fig:qua-qrw-scheme2}(c), respectively.  As shown in the plots, one may observe two distinct behaviours $\sim m^{-3}$ (dashed line) and $\sim m^{-5/2}$ (dash-dotted line).}
\label{fig:qrw-f-vs-m-exp-pow-typ}
\end{figure}

So far we have discussed the behavior of the average quantities, but even the typical survival probability $S^{\star}_m$ and the typical first-detection probability $F^{\star}_m$ behave in a manner similar to their respective averages.  Some representative plots are given in Fig.~\ref{fig:qua-qrw-scheme2},  panels (d) -- (f), and in Fig.~\ref{fig:qrw-f-vs-m-exp-pow-typ}.

\begin{figure}[!htbp]
\centering
\includegraphics[scale=1]{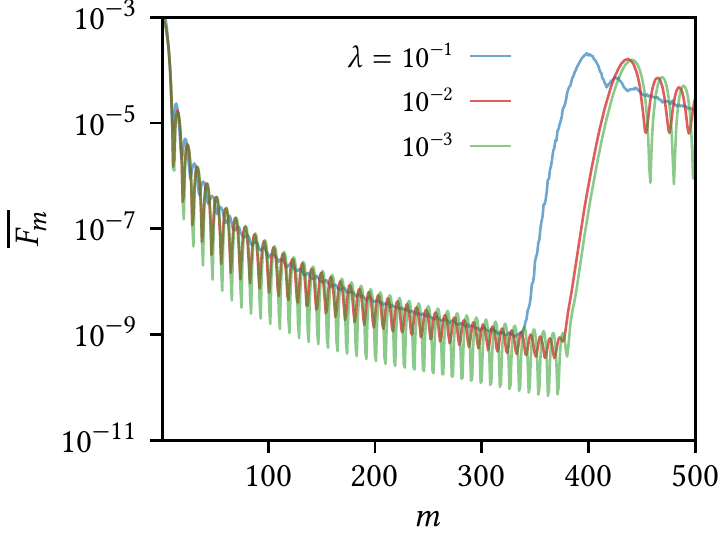}
\caption{Average first-detection probability for the QRW model (\textbf{Scheme 2}) for the Poisson distribution~\eqref{eq:qua-Poisson} with different values of $\lambda$.  Here, the parameters $n_0$,  $a$,  $b$,  and $\theta$ have the same values as in Fig.~\ref{fig:qua-qrw-scheme2}.  The system size is $N=150$, while the number of averaging realizations is $1000$. }
\label{fig:qrw-fm-osc}
\end{figure}

Before closing this part, the last issue that we discuss is the behavior of $\overline{S_m}$ and $\overline{F_m}$ for small $m \ll m_1^\star(N)$.  As it turns out, these quantities show oscillations in such a regime, with the oscillations becoming more pronounced as the distribution $p_\tau$ becomes more narrow.  To show a representative result,  we plot in Fig.~\ref{fig:qrw-fm-osc} the quantity $\overline{F_m}$ versus $m$ for the case of the Poisson distribution~\eqref{eq:qua-Poisson} as one varies the parameter $\lambda$ that controls the narrowness of the distribution. It is evident from the plot that with decrease of $\lambda$,  oscillations in the behavior of $\overline{F_m}$ become more pronounced. This feature is generic across the different $p_\tau$'s that we considered (of course, for the delta-function distribution~\eqref{eq:qua-delta}, the oscillations look most striking).

On the basis of the foregoing, we see a stark contrast between our
obtained results on the survival probability under the two choices of
the dynamics. It is our objective in the following to offer an
analytical treatment of these results. 

\subsubsection{Analytical results}
\label{sec:QRW3b}

In \textbf{Scheme 1}, we obtain $|\psi_m^{(\rm a)}\rangle$ for arbitrary $m$ as
\begin{align}
        \ket{\psi_m^{({\rm a})}}=PU_m \dots PU_3PU_2PU_1\i.
\label{eq:qua-psim-definition-1}
\end{align}
From Eqs.~\eqref{eq:qua-Sm-definition},~\eqref{eq:qua-P-operator-def},
and~\eqref{eq:qua-psim-definition-1}, we see that
$S_m$ may be expressed as
\begin{align}
S_m = \prod_{\alpha=1}^{m} ~ q(\tau_\alpha), 
        \label{eq:qua-Sm-1}
\end{align}
where the (quantum) probability $q(\tau_\alpha)$ is given by
\begin{align}
q(\tau_\alpha) &\equiv \left|\langle\psi(0)|U_\alpha|\psi(0)\rangle\right|^{2}  \label{eq:qua-qrw-qtau-0} \\
          &= \frac{1}{|a|^2 + |b|^2}\left| a^{*} \Psi_{\rm
          u}(n_0,\tau_\alpha) + b^{*} \Psi_{\rm d}(n_0,\tau_\alpha) \right|^2,
        \label{eq:qua-qrw-qtau}
\end{align}
with $\Psi_{\rm u}(n_0,\tau_\alpha)$ and $\Psi_{\rm d}(n_0,\tau_\alpha)$ given by
Eqs.~\eqref{eq:qua-psiud-final} and~\eqref{eq:qua-psiud-even-final}.
Note that $q(\tau)$ is nothing but the probability for the walker to be found in the
initial state $|\psi(0)\rangle$ after evolution for time $\tau$.
Using Eq.~\eqref{eq:qua-Sm-1}, one obtains the average survival probability as
\begin{align}
\overline{S_m}=\prod_{i=1}^m \left( \sum_\tau p_\tau ~ q(\tau)\right) = \exp\left( m \log \sum_\tau p_\tau ~ q(\tau) \right).
        \label{eq:qua-Sm-avg-qrw}
\end{align}
As shown in~\ref{sec:qua-app1}, the most probable value of the 
survival probability $S_m^\star$ is obtained as
\begin{align}
S_m^\star=\exp\left( m \sum_\tau p_\tau ~ \log q(\tau) \right).
        \label{eq:qua-Sm-typical-qrw}
\end{align}
Using the Jensen's inequality
$\overline{\exp
(x)} \ge \exp(\,\overline{x}\,)$, we obtain that
\begin{align}
        \overline{S_m} \ge S_m^\star,
\end{align}
with the equality holding only when there is no randomness in $\tau$,
that is, when $\tau$ can take on only a single value:
$p_\tau=\delta_{\tau,\tau_0}$, with $\tau_0 >0$. On performing a large number $m$ of projective
measurements, the value of the survival probability to remain in the initial state that is measured in a single
experimental run will equal $S_m^\star$ in the limit $m\to \infty$. On
the other hand, averaging the survival probability over a large (ideally
infinite) number of experimental runs would yield the value $\overline{S_m}$.

The theoretical results are compared against those obtained in 
numerical implementation of the \textbf{Scheme 1} dynamics in
Fig.~\ref{fig:qua-qrw-scheme1}. In this figure, the continuous lines in
red obtained from Eq.~\eqref{eq:qua-Sm-avg-qrw} show the behavior of
$\overline{S_m}$, while the blue-dashed lines obtained from Eq.~\eqref{eq:qua-Sm-typical-qrw} depict that of $S^{\star}_m$. 
We see from the figure a very good match between theoretical and numerical results for the average survival probability, whereas the numerical results for the most probable survival probability  fluctuate around the theoretical estimate as expected.

In \textbf{Scheme 2}, we obtain $\ket{\psi_m^{({\rm a})}}$ for arbitrary $m$ as
\begin{align}
        \ket{\psi_m^{({\rm a})}}=\widetilde{P} U_m \ldots  \widetilde{P} U_3 \widetilde{P} U_2 \widetilde{P} U_1\i.
\label{eq:qua-psim-definition-2}
\end{align}
In this scheme, to find an analytical closed form of $\overline{S_m}$
and $S_m^\star$ is
non-trivial, as we explain below. Consequently, we rely on a semi-analytical approach that involves implementation of the following four steps:
\begin{enumerate}
        \item For a given choice of $a,~b,$ and $n_0$ specifying the initial
state $|\psi(0)\rangle$ in Eq.~\eqref{eq:qua-psi0}, we first obtain the vector $\ket{\Psi(n,0)}$ by using Eqs.~\eqref{eq:qua-psi-n} and~\eqref{eq:qrw-two-comp-vector}. Then, we implement the discrete Fourier transform given by 
$\ket{\tilde\Psi(k,0)} =
\sum_{n}
\ket{\Psi(n,0)} ~ \ee^{\ii 2\pi kn/N} $.
\item Subsequently, we use Eq.~\eqref{eq:qua-iteration} to obtain 
$\ket{\tilde\Psi(k,\tau_1)}$ as the result of dynamical evolution for a random time $\tau_1$ sampled
according to either the exponential
distribution~\eqref{eq:qua-app-exp} or the power-law
distribution~\eqref{eq:qua-app-pow}, and with $\ket{\tilde\Psi(k,0)}$ as
                the initial condition. At the end of the evolution, one has
the set $\{\ket{\tilde\Psi(k,\tau_1)}\}$.
\item Next, we perform inverse discrete Fourier transform of the set
$\{\ket{\tilde\Psi(k,\tau_1)}\}$ to obtain the set $\{\ket{\Psi(n,\tau_1)}\}$. To
implement a projective measurement at the end of evolution for time
$\tau_1$ and obtaining the corresponding leftover component of the state, we first
obtain the state $|\psi(\tau_1)\ra$ by using the obtained values of
the elements of $\{|\Psi(n,\tau_1)\ra \}$ in
Eq.~\eqref{eq:qua-psi-n}, and then compute the
                difference $|\psi(\tau_1)\ra-|\psi(0)\ra\la
                \psi(0)|\psi(\tau_1)\ra$, which yields the state
                $|\psi_1^{({\rm a})}\ra$. 
\item Steps 1--3 are applied in turn to the leftover component of the 
state corresponding to last projection; $m \ge 1$ number of repetitions would 
generate the leftover component of the instantaneous state after $m$ projections, and this allows us
to obtain the survival probability $S_m$ for a given realization
$\{\tau_\alpha\}_{1\le \alpha \le m}$ of the dynamics.
\end{enumerate}
The method is semi-analytical in the sense that while the dynamical
evolution in the Fourier space follows the exact
solution~\eqref{eq:qua-iteration}, inverse Fourier transform to obtain
$|\Psi(n,\tau)\ra$ is performed numerically. 
The semi-analytical results for the average and the typical survival
probability are compared in Fig.~\ref{fig:qua-qrw-scheme2} against
numerical results, demonstrating a very good match. 

We remark that in {\bf Scheme 1}, each time a projective measurement is
made, the dynamical evolution starts afresh from a state that is just the initial state multiplied by a complex number
(e.g.,  the state for subsequent dynamical evolution after say the
first measurement is $|\psi^{({\rm a})}_1\ra= \la \psi(0)|\psi(\tau_1)\ra |\psi(0)\ra$).
Consequently,  the
survival probability $S_m$ is a product of the quantum probabilities 
$q(\tau_\alpha)$ over different times $\tau_\alpha$, see
Eq.~\eqref{eq:qua-Sm-1}.  This is however not the case in {\bf Scheme 2},  and this defies the survival probability $S_m$
to be written as a product of probabilities for different
$\tau_\alpha$'s and consequently, a straightforward analytical estimate
of the average and the typical survival probability for {\bf Scheme 2}.

\section{Tight-binding model (TBM)}
\label{sec:TBM}

\subsection{Model and site occupation probability}
\label{sec:TBM1}

The tight-binding model (TBM) that we now study involves quantum evolution of a particle
on a one-dimensional lattice that we consider here to be of $N$ sites
with periodic boundary conditions. The dynamics in continuous time is generated by the Hamiltonian~\cite{qua-Dunlap:1986,qua-Dunlap:1988}
\begin{align}
H=-\gamma \sum_{j=0}^{N-1} (|j+1\rangle \langle j|+|j\rangle \langle
j+1|) \, ; \quad |N\rangle=|0\rangle.
\label{eq:qua-H}
\end{align}
Here, $\gamma >0$ is a real parameter, while the index $j$ denotes the lattice sites. Let
$\psi_{n,n_0}(t)= \bk{n}{\psi(t)}$, with $n=0,1,2,\ldots,N-1$
be the probability amplitude to find the particle on site $n$ at time
$t$ while starting from site $n_0$ at time $t=0$, with the normalization
$\sum_{n=0}^{N-1} |\psi_{n,n_0}(t)|^2=1~\forall~t$. From the evolution
equation $| \psi(t+\Delta t) \ra = \exp(-\ii H \Delta t) | \psi(t) \ra; \quad | \psi(0) \ra = | n_0 \ra$, one obtains the time
evolution of $\psi_{n,n_0}(t)$ in a small time interval $\Delta t$  as 
\begin{align}
\psi_{n,n_0}(t+\Delta t)=\sum_{j=0}^{N-1} ~\langle n |\ee^{-\ii H\Delta t} |j\rangle ~ \psi_{j,n_0}(t).
\end{align}
Expanding in powers of $\Delta t$ the right hand side of the
above equation and then taking the limit of
continuous time, $\Delta t \to 0$, one obtains the evolution
$\partial \psi_{n,n_0}(t)/\partial t=-\ii\sum_{j=0}^{N-1} \langle
n|H|j\rangle \psi_{j,n_0}(t)$. Equation~\eqref{eq:qua-H} gives $\langle
j|H|k\rangle=-\gamma~(\delta_{j,k-1}+\delta_{j,k+1})$, yielding
\begin{align}
\frac{\partial \psi_{n,n_0}(t)}{\partial t}=\ii\gamma~(
\psi_{n-1,n_0}(t)+\psi_{n+1,n_0}(t)).
\label{eq:qua-psin-time-evolution-1}
\end{align}

\begin{figure}[!htbp]
\centering
\includegraphics[scale=0.85]{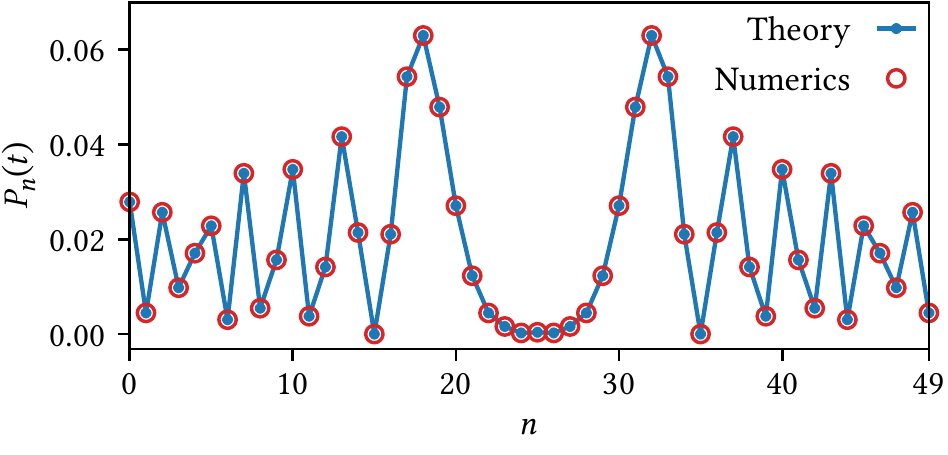}
        \caption{Site occupation probability $P_n(t)$ in the TBM on
a one-dimensional periodic lattice of $N$ sites and at time $t=10$, while starting
from initial site $n_0=0$. Here, we have 
$N=50, \gamma=1$. Numerics
correspond to bare evolution of the unitary dynamics of the model, while
analytical results are given by Eq.~\eqref{eq:qua-pn}. The line is a guide to the eye.}
\label{fig:qua-tbm-site-probability}
\end{figure}

In order to solve Eq.~\eqref{eq:qua-psin-time-evolution-1} for $\psi_{n,n_0}(t)$, we perform
discrete Fourier transform of the set $\{\psi_{n,n_0}(t)\}_{0 \le n \le N-1}$, given by the set
$\{\widehat{\psi}_{q|n_0}(t)\}_{ 0 \le q \le N-1}$, with
$\widehat{\psi}_{q|n_0}(t)=\sum_{j=0}^{N-1}\psi_{j,n_0}(t) \exp(-\ii 2\pi
jq/N); \quad
\psi_{j,n_0}(t)=(1/N)\sum_{q=0}^{N-1}\widehat{\psi}_{q|n_0}(t)\exp(\ii 2\pi
jq/N)$. From Eq.~\eqref{eq:qua-psin-time-evolution-1}, one then obtains
\begin{align}
\frac{\partial \widehat{\psi}_{q|n_0}(t)}{\partial
t}=2\ii\gamma\cos\left(\frac{2\pi
q}{N}\right)\widehat{\psi}_{q|n_0}(t).
\label{eq:qua-psiq-equation}
\end{align}
Subject to the initial condition $\psi_{n,n_0}(0)=\delta_{n,n_0}$
implying $\widehat{\psi}_{q|n_0}(0)=\exp(-\ii 2\pi
        n_0q/N)$, Eq.~\eqref{eq:qua-psiq-equation} has the solution
        $\widehat{\psi}_{q|n_0}(t)=\exp(\ii 2 \gamma t \cos (2\pi
        q/N)-\ii 2\pi n_0 q/N)$,
inverting which yields
\begin{align}
\psi_{j,n_0}(t)=\frac{1}{N}\sum_{q=0}^{N-1}\ee^{\ii 2\gamma t \cos (2\pi q/N)+\ii 2\pi
        q(j-n_0)/N}.
        \label{eq:qua-solution-psij}
\end{align}
It is easily checked that $\sum_{j=0}^{N-1}|\psi_{j,n_0}(t)|^2=1$, as
required. In particular, starting with the particle on site $n_0$, we may ask for
the probability $P_n(t)$ to be on site $n$ at time $t$, the so-called site
occupation probability. It is given by
\begin{align}
        P_n(t)=|\psi_{n,n_0}(t)|^2=\left|\frac{1}{N}\sum_{q=0}^{N-1} \ee^{\ii 2 \gamma t \cos (2\pi q/N)+\ii 2\pi
        q(n-n_0)/N}\right|^2.
\label{eq:qua-pn}
\end{align}
In Fig.~\ref{fig:qua-tbm-site-probability}, we show a comparison between
numerical results and theory, demonstrating a perfect match.
In the figure, numerics
correspond to bare evolution of the unitary dynamics of the model, while
analytical results are given by Eq.~\eqref{eq:qua-pn}.

\subsection{Random projective measurements and survival probability}
\label{sec:TBM2}

\subsubsection{Numerical results}
\label{sec:TBM2a}

Here, we report numerical results on the average and the typical
survival probability for the TBM subject to projective measurements at
random times. A typical time evolution of the system is shown in
Fig.~\ref{fig:qua-evolution-schematic}. The initial state corresponds to the particle located on
site $n_0$ (thus,
$|\psi(0)\rangle=|n_0\rangle$), and as in the case of the QRW reported
above, we consider the projective measurement to be involving projection
to the initial state $n_0$ implemented by the projection
operator $P = |n_0\rangle \langle n_0|$.  Here, the random variable $\tau$
between two successive measurements, now a continuous variable, is taken to be
distributed according to either an exponential distribution:
\begin{align}
p(\tau)=r\exp(-r\tau) \,; \quad \tau \in [0,\infty), ~r>0 \, ; \quad \langle \tau \rangle = \frac{1}{r} \, , \quad {\mathrm{Var}}[\tau] = \frac{1}{r^2} \, ,
\label{eq:qua-exponential-ptau}
\end{align}
or a power-law distribution:
\begin{equation}
\begin{aligned}
& p(\tau)=\frac{\alpha}{\tau_{\rm ch}(\tau/\tau_{\rm ch})^{1+\alpha}}  \, ; \quad \tau \in [\tau_{\rm ch},\infty) \,, ~\alpha > 0  \, ;   \\
& \langle \tau \rangle = \frac{ \tau_{\rm ch} \, \alpha}{\alpha - 1} \,  , ~\alpha > 1 \,; \quad {\mathrm{Var}}[\tau] = \frac{\tau^2_{\mathrm{ch}} \alpha }{(\alpha-2) (\alpha-1)^2}\, , ~\alpha >2\, .
\label{eq:qua-powerlaw-ptau}
\end{aligned}
\end{equation}
In Eq.~\eqref{eq:qua-powerlaw-ptau}, the parameter $\tau_{\rm
ch}>0$ sets the lower cut-off scale. The exponential distribution~\eqref{eq:qua-exponential-ptau} has a finite mean for all
values of $r$. On the other hand, the power-law distribution~\eqref{eq:qua-powerlaw-ptau} has
a finite mean only for $\alpha>1$ and a finite variance only for $\alpha>2$, and since we would like in the view
of discussions presented in Section~\ref{sec:QRW3} to have the
$\tau_\alpha$'s to have a finite average, we will in the following
consider values of $\alpha$ to be larger than $1$.   

\begin{figure}[!htbp]
\centering
\includegraphics[scale=0.85]{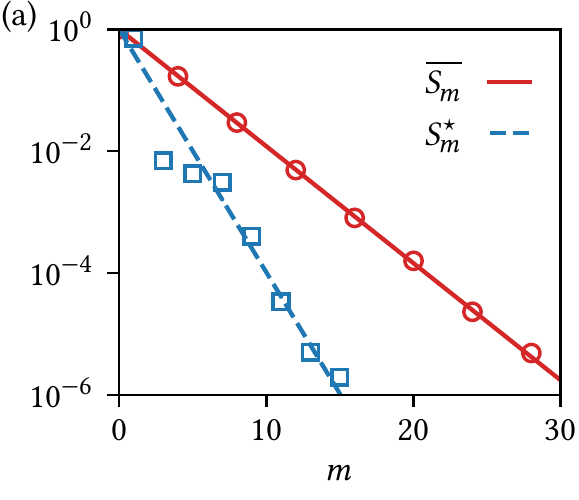}
\includegraphics[scale=0.85]{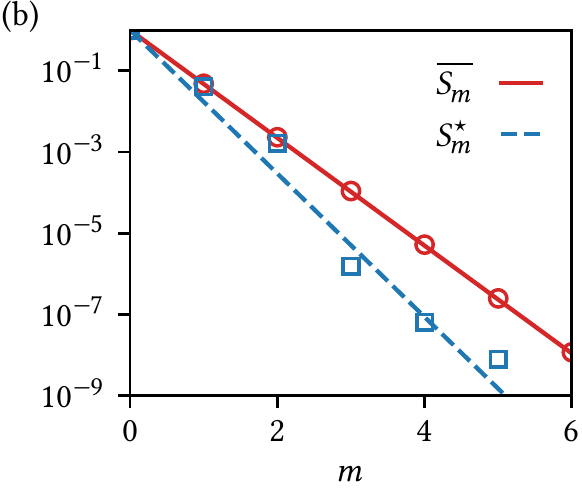} 
\includegraphics[scale=0.85]{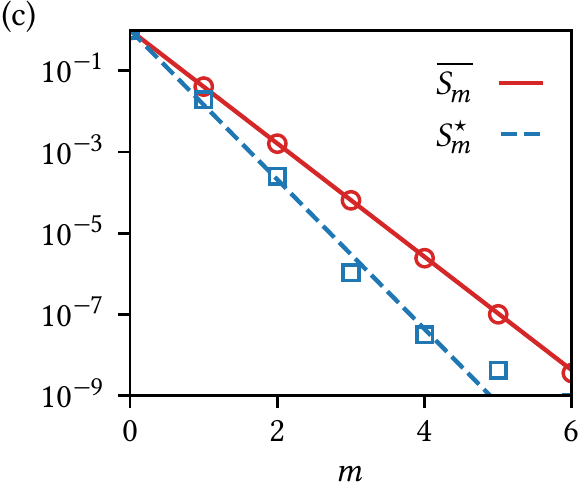}
\caption{Average and typical survival probability for the TBM subject to instantaneous projective measurements at random times
(\textbf{Scheme 1}). The plots correspond to the initial state $|0\ra$
        (particle located on site $n_0=0$) that is subject to repeated projective
measurements at random times to the initial state and subsequent evolution with the
projected component of the instantaneous state. Here, the time intervals $\tau_\alpha$ between two consecutive
measurements are i.i.d.~random variables sampled from the exponential
distribution~\eqref{eq:qua-exponential-ptau} with $r=2.0$ (panel (a)) and
from the power-law distribution~\eqref{eq:qua-powerlaw-ptau} with $\tau_{\rm ch} = 1$ and
$\alpha=2.5$ (panel (b)) and $\alpha=3.5$ (panel (c)). The system size
        is $N=200$, while we have taken $\gamma=1$. In the plots, the points are based on results
        obtained from numerical implementation of the dynamics; while
        the average survival probability $\overline{S_m}$ involves
        averaging $10^4$ realizations of the set $\{\tau_\alpha\}_{1\le
        \alpha \le m}$, the typical survival probability $S_m^\star$
        corresponds to results obtained in a typical realization of the
        $\tau_\alpha$'s. The lines in the plots correspond to analytical
        results given by
        Eqs.~\eqref{eq:survival-prob-TBM-projected} and~\eqref{eq:qua-tbm-qtau}.}
\label{fig:qua-tbm-proj}
\end{figure}

Figure~\ref{fig:qua-tbm-proj} shows in case of the evolution
under \textbf{Scheme 1} our numerical results on the average survival
probability $\overline{S_m}$ and the typical survival probability $S_m^\star$, both plotted as a
function the $m$, the number of measurements.  In the figure, panel (a)
corresponds to the exponential distribution~\eqref{eq:qua-exponential-ptau} for the $\tau$ with $r=2$, while panels (b) and (c) are for
the power-law distribution~\eqref{eq:qua-powerlaw-ptau} with $\alpha$ having values
$2.5$ and $3.5$, respectively. The panels suggest for both
the quantities $\overline{S_m}$ and $S_m^\star$ an
exponential decay with $m$ for large $m$. 

\begin{figure}[!htbp]
\centering
\includegraphics[scale=0.75]{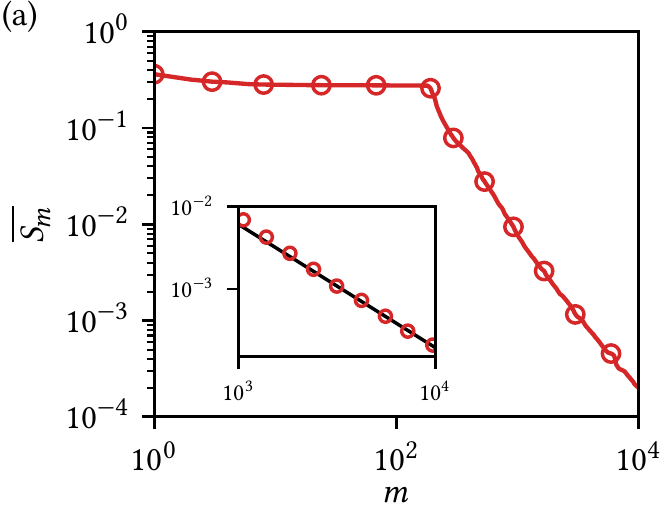} \hskip1pt 
\includegraphics[scale=0.75]{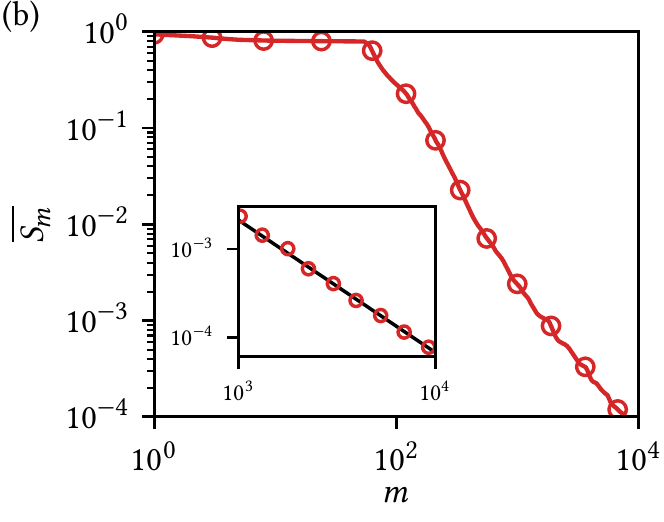}  \hskip1pt 
\includegraphics[scale=0.75]{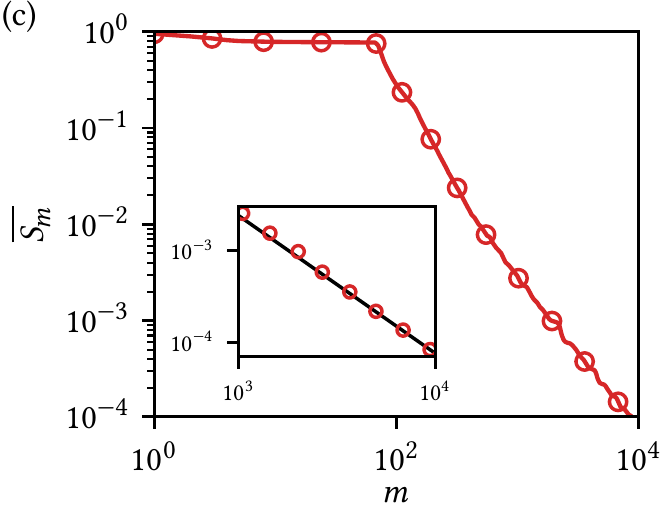}  \\[2ex]
\includegraphics[scale=0.76]{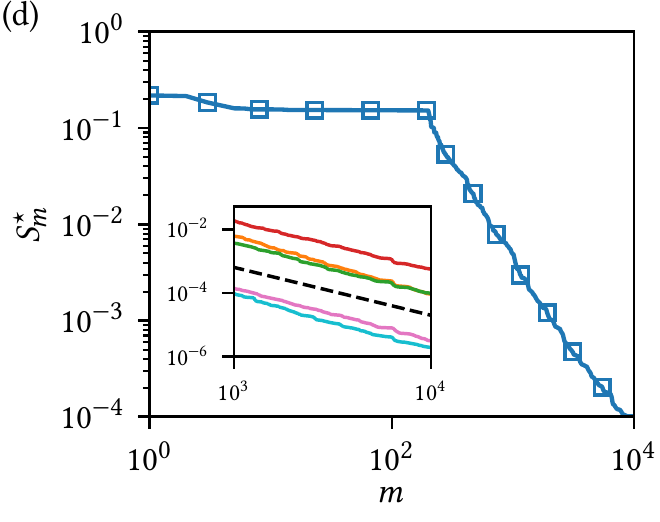}    \hskip1pt 
\includegraphics[scale=0.76]{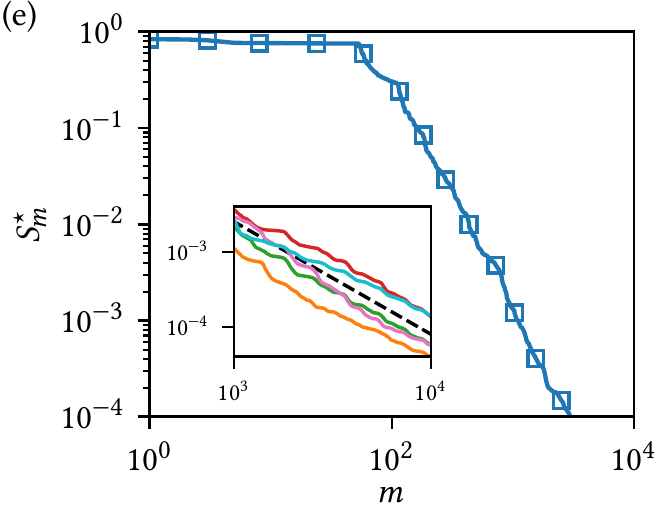}   \hskip1pt  
\includegraphics[scale=0.76]{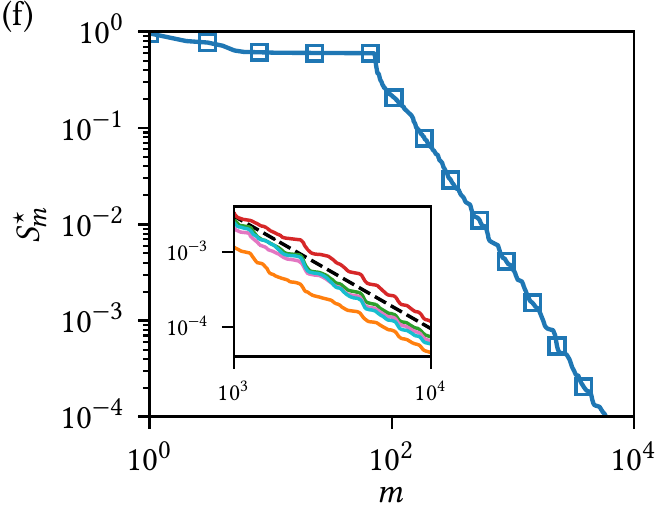}
\caption{Average survival probability (panels (a) -- (c)) and
        typical survival probability (panels (d) -- (f)) for the TBM subject to instantaneous projective measurements at random times
(\textbf{Scheme 2}). The plots correspond to the initial state $|0\ra$
        (particle located on site $n_0=0$) that is subject to repeated projective
measurements at random times to the initial state and subsequent
        evolution with the leftover component of the
        instantaneous state after the measurement. Here, the time intervals $\tau_\alpha$ between two consecutive
measurements are i.i.d.~random variables sampled from the exponential
distribution~\eqref{eq:qua-exponential-ptau} with $r=2$ (panels (a) and
        (d)), and from the power-law distribution~\eqref{eq:qua-powerlaw-ptau} with $\tau_{\rm ch} = 1$ and 
        $\alpha=2.5$ (panels (b) and (e)) and $\alpha=3.5$ (panels (c) and (f)). The system size
        is $N=200$, while we have taken $\gamma=1$. In the main plots, the points are based on results
        obtained from numerical implementation of the dynamics; while
        the average survival probability $\overline{S_m}$ involves
        averaging over $25$ realizations of the set $\{\tau_\alpha\}_{1\le
        \alpha \le m}$, the typical survival probability $S_m^\star$
        corresponds to results obtained in a typical realization of the
        $\tau_\alpha$'s. The lines in the main plots correspond to those
        obtained from the semi-analytical approach described in the
        text, see Section~\ref{sec:TBM2b}. In the insets in the upper
        row, the points correspond to numerically-evaluated average
        survival probability, while the line represents an $m^{-3/2}$
        behavior. In the insets in the lower
        row, the continuous lines correspond to numerically-evaluated
        survival probability for five typical realizations of
        $\{\tau_\alpha\}_{1\le \alpha \le m}$, while the dashed line
        represents an $m^{-3/2}$ behavior. We conclude from the insets
        in both the upper and the lower row that the average as well as the typical
        survival probability behaves at large $m$ as $m^{-3/2}$.} 
\label{fig:qua-tbm-left}
\end{figure}

\begin{figure}[!htbp]
\centering
\includegraphics[scale=1]{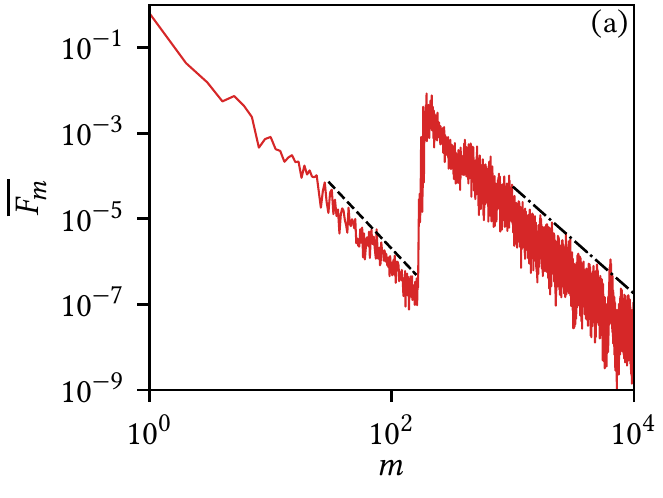} \hskip5pt 
\includegraphics[scale=1]{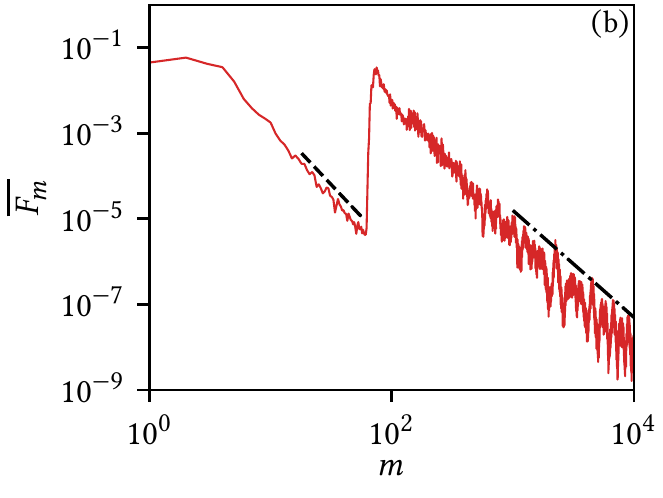}   
\caption{Average first-detection probability for the TBM (\textbf{Scheme 2}).  For panels (a) and (b), parameter values and number of averaging realizations are the same as in Figs.~\ref{fig:qua-tbm-left}(a) and~\ref{fig:qua-tbm-left}(c), respectively.  As shown in the plots, one may observe two distinct behaviours $\sim m^{-3}$ (dashed line) and $\sim m^{-5/2}$ (dash-dotted line).}
\label{fig:tbm-f-vs-m-exp-pow}
\end{figure}

\begin{figure}[!htbp]
\centering
\includegraphics[scale=1]{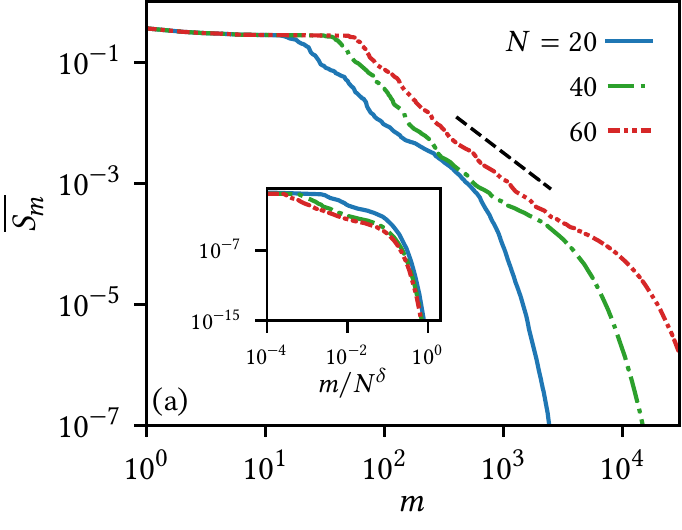} \hskip1pt 
\includegraphics[scale=1]{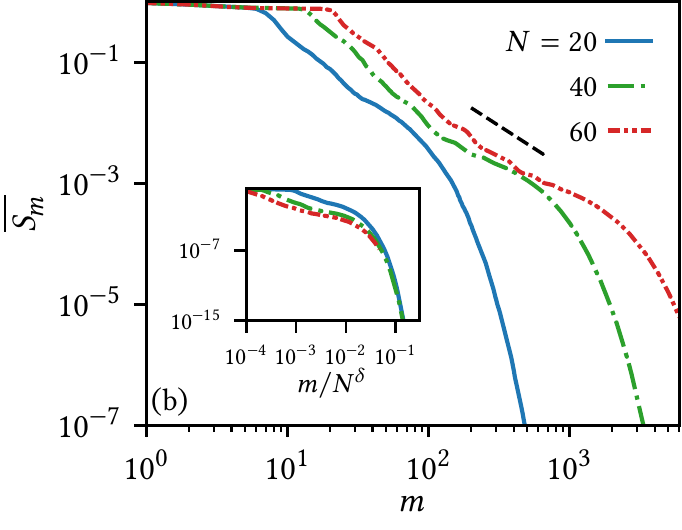}   
\caption{Average survival probability for the TBM (\textbf{Scheme 2}).  For panels (a) and (b), parameter values other than $N$ and the number of averaging realizations are the same as in Figs.~\ref{fig:qua-tbm-left}(a) and~\ref{fig:qua-tbm-left}(c), respectively.  The main plots show a crossover from a $m^{-3/2}$-behaviour (dashed line) to an exponential tail over the characteristic value $m_2^\star(N)$ of $m$; the collapse of the data shown in the insets suggests the scaling~\eqref{eq:m2-scaling-QRW}.}
\label{fig:tbm-N-dependence}
\end{figure}

\begin{figure}[!htbp]
\centering
\includegraphics[scale=1]{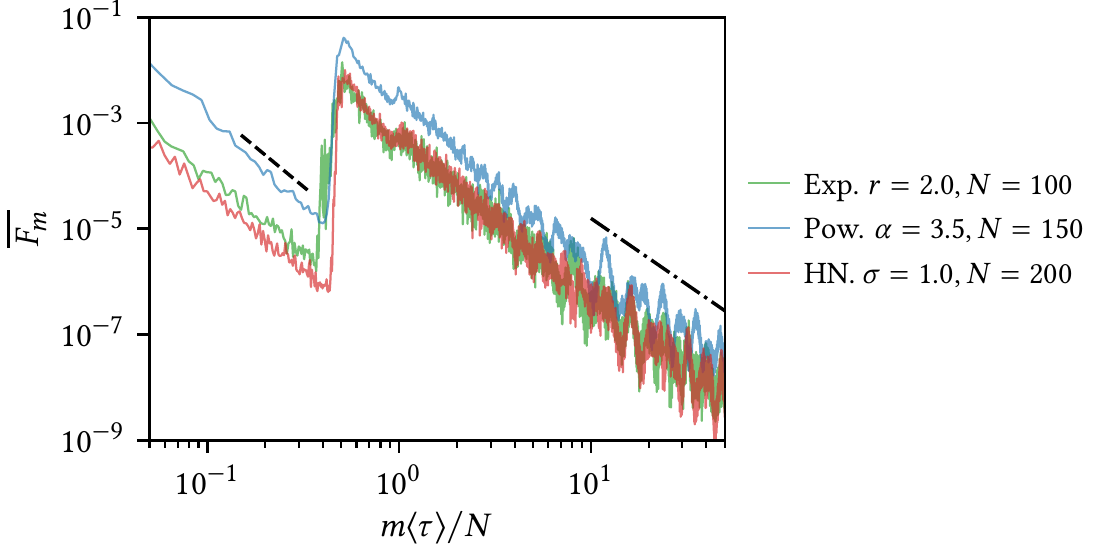}
\caption{Average first-detection probability for the TBM (\textbf{Scheme 2}) for the exponential distribution~(\ref{eq:qua-exponential-ptau}), the power-law distribution~(\ref{eq:qua-powerlaw-ptau}), and the half-normal distribution~(\ref{eq:qua-hn-ptau}).  Here, the parameters $n_0$ and $\gamma$ have the same values as in Fig.~\ref{fig:qua-tbm-left}.  The number of averaging realizations in all cases is $50$.  The parameter $r$ for the exponential distribution and the parameters $\alpha$ and $\tau_{\rm ch}$ for the power-law distribution have the same values as in Fig.~\ref{fig:qua-tbm-left}, panels (a) and (c), respectively.  
For the data plotted for the half-normal distribution~(\ref{eq:qua-hn-ptau}), we have $\tau_{\mathrm{hn}} = 0.0$ and $\sigma=1.0$.
The plot suggests the scaling depicted in Eq.~(\ref{eq:m1-scaling-QRW}). }
\label{fig:tbm-fm}
\end{figure}

\begin{figure}[!htbp]
\centering
\includegraphics[scale=1.1]{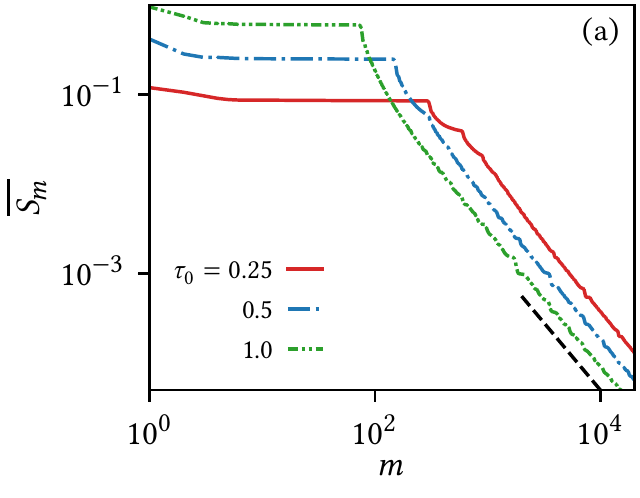} \hskip5pt 
\includegraphics[scale=1.1]{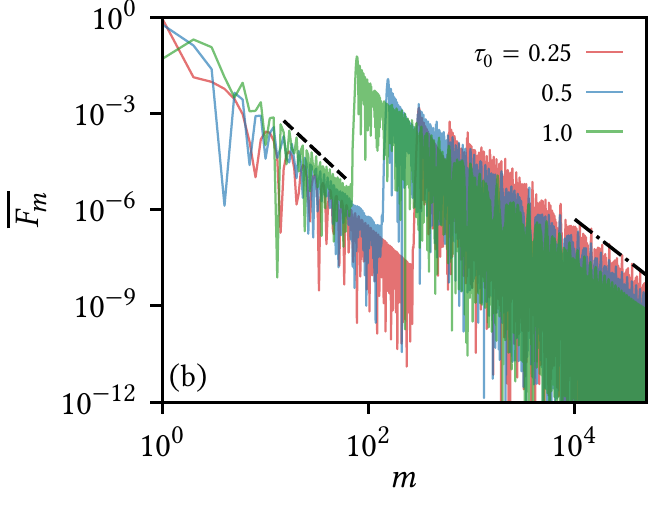}   
\caption{Average survival and first-detection probability for the TBM (\textbf{Scheme 2}).  Here, the parameters $n_0$ and $\gamma$ have the same values as in Fig.~\ref{fig:qua-tbm-left}.  Here, the time intervals $\tau_\alpha$ between two consecutive
measurements are i.i.d.~random variables sampled from the
        distribution~(\ref{eq:qua-delta-ptau}).  The values of the parameter $\tau_0$ are shown in the figure.  The system size
        is $N=150$.  In panel (a),  as shown in the plot, one may observe a behavior $\sim m^{-3/2}$ (dashed line), while in panel (b),  two distinct behaviours $\sim m^{-3}$ (dashed line) and $\sim m^{-5/2}$ (dash-dotted line) may be seen.  Note that since $\tau$ has only one allowed value, namely, $\tau_0$,  it is redundant to use the overbar over $S_m$ and $F_m$ to denote their average values.}
\label{fig:tbm-samet-sp-fp}
\end{figure}

\begin{figure}[!htbp]
\centering
\includegraphics[scale=1.1]{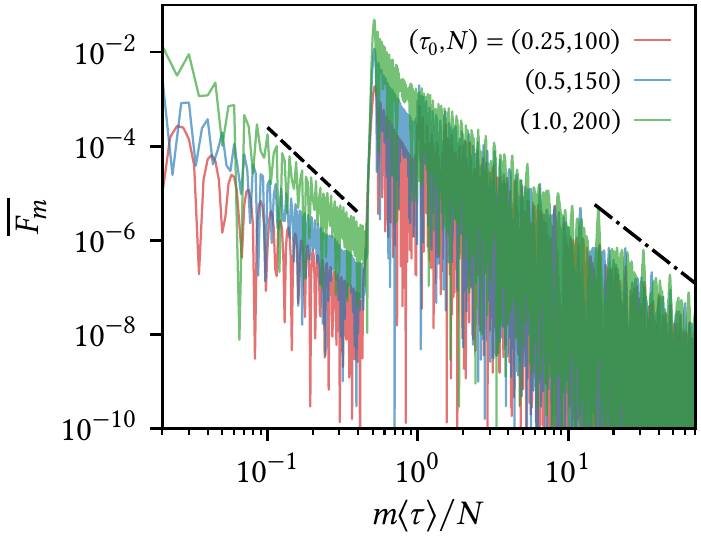}   
\caption{Average first-detection probability for the TBM (\textbf{Scheme 2}). Parameter values and other details are the same as in Fig.~\ref{fig:tbm-samet-sp-fp}.  The plot suggests that the crossover from the $\sim m^{-3}$-behaviour (dashed line) to the $\sim m^{-5/2}$-behavior (dash-dotted line) over the characteristic value $m_1^\star(N)$ of $m$ that follows the scaling~\eqref{eq:m1-scaling-QRW}.  Note that since $\tau$ has only one allowed value, namely, $\tau_0$,  it is redundant to use the overbar over $F_m$ to denote its average.}
\label{fig:tbm-fp-scaled}
\end{figure}

\begin{figure}[!htbp]
\centering
\includegraphics[scale=1]{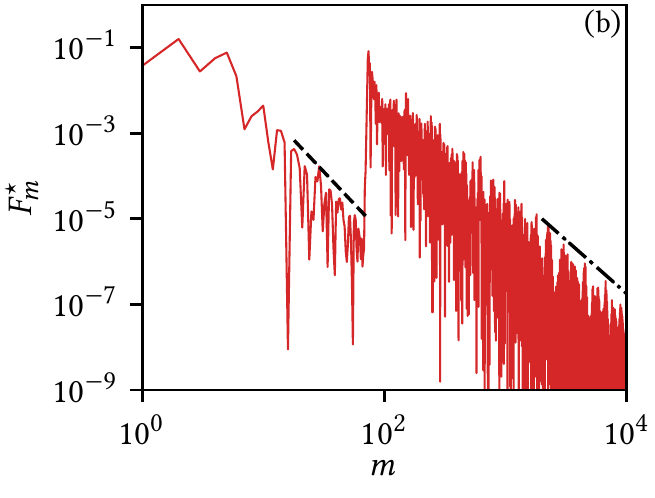} \hskip5pt 
\includegraphics[scale=1]{plot-TBM-fm-vs-m-leftover-direct-n200-i100-p100-m1e4-g1-nh25-pow3-5-typ.pdf}   
\caption{Typical first-detection probability for the TBM (\textbf{Scheme 2}).  For panels (a) and (b), parameter values are the same as in Figs.~\ref{fig:qua-tbm-left}(a) and~\ref{fig:qua-tbm-left}(c), respectively.  As shown in the plots, one may observe two distinct behaviours $\sim m^{-3}$ (dashed line) and $\sim m^{-5/2}$ (dash-dotted line).}
\label{fig:tbm-f-vs-m-exp-pow-typ}
\end{figure}

\begin{figure}[!htbp]
\centering
\includegraphics[scale=1]{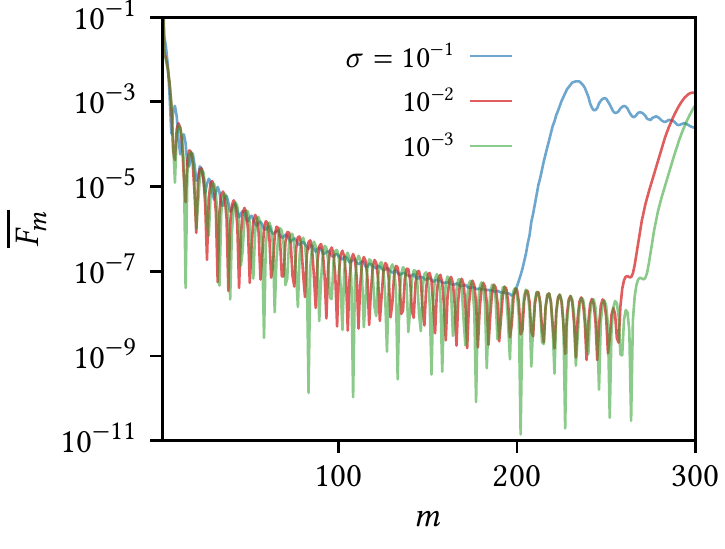}
\caption{Average first-detection probability for the TBM (\textbf{Scheme 2}) for the half-normal distribution~\eqref{eq:qua-hn-ptau} with $\tau_{\mathrm{hn}} = 0.25$ and different values of $\sigma$.  Here, the parameters $n_0$ and $\gamma$ have the same values as in Fig.~\ref{fig:qua-tbm-left}.  The system size is $N=150$, while the number of averaging realizations is $500$. }
\label{fig:tbm-fm-osc}
\end{figure}

As regards {\bf Scheme 2} dynamics,  the behavior of both the average and the typical value of the survival and the first-detection probability is identical to what we reported for the QRW, see Section~\ref{sec:QRW3a}.  The statement holds not just for the exponential distribution~(\ref{eq:qua-exponential-ptau}) and the power-law distribution~(\ref{eq:qua-powerlaw-ptau}), but also for the delta-function distribution corresponding to measurements performed at regular intervals:
\begin{align}
p(\tau)=\delta(\tau-\tau_0);~\quad \tau_0>0\, ; \quad \langle \tau \rangle = \tau_0 \, , \quad {\mathrm{Var}}[\tau] = 0 \, ,
\label{eq:qua-delta-ptau}
\end{align}
and for the half-normal distribution:
\begin{equation}
\begin{aligned}
& p(\tau)= \sqrt{\frac{2}{ \pi \sigma^2} } \, \exp\left[ - \frac{(\tau-\tau_{\mathrm{hn}})^2}{2 \sigma^2} \right]  ; \quad \tau \in [\tau_{\mathrm{hn}},\infty) \,, ~\sigma > 0  \, ; \\ 
& \langle \tau \rangle = \tau_{\mathrm{hn}} + \sigma \sqrt{\frac{2}{\pi}} \,  , \quad {\mathrm{Var}}[\tau] = \sigma^2\left(1 -\frac{2}{\pi}\right)\, ,
\label{eq:qua-hn-ptau}
\end{aligned}
\end{equation}
in which the quantity $\tau_{\rm hn}$ sets the lower limit of $\tau$.
Indeed, as shown in Figs.~\ref{fig:qua-tbm-left} and~\ref{fig:tbm-f-vs-m-exp-pow}, one observes in $\overline{S_m}$ and $\overline{F_m}$ the occurrence of the $m^{-3/2}$ behaviour in the former and of $m^{-3}$ and $m^{-5/2}$ behavior in the latter.  The crossover between the $m^{-3}$ and the $m^{-5/2}$ behavior takes place over the characteristic value $m_1^\star(N)$ that satisfies the scaling~\eqref{eq:m1-scaling-QRW}, as implied by the results in Fig.~\ref{fig:tbm-fm}.  The existence of the characteristic value $m_2^\star(N)$ satisfying the scaling~\eqref{eq:m2-scaling-QRW} may be seen in Fig.~\ref{fig:tbm-N-dependence}.  The results for the delta-function distribution~\eqref{eq:qua-delta-ptau} are included in Figs.~\ref{fig:tbm-samet-sp-fp} and \ref{fig:tbm-fp-scaled}, while those for the typical survival probability in the case of the exponential distribution~\eqref{eq:qua-exponential-ptau} and \eqref{eq:qua-powerlaw-ptau} are shown in Fig.~\ref{fig:tbm-f-vs-m-exp-pow-typ}.  The existence of oscillations in the behavior of $\overline{F_m}$ at small $m < m_1^\star(N)$ that become more pronounced as the distribution $p(\tau)$ of the time interval between successive measurements is shown for the case of the half-normal distribution~\eqref{eq:qua-hn-ptau} in Fig.~\ref{fig:tbm-fm-osc}.  Let us remark that our results summarized above hold for finite $N$,  while if one studies an infinite system (i.e., the limit $N \to\infty$), the scale $m_1^\star(N)$ diverges, and one observes the large-$m$ behavior $\overline{S_m} \sim m^{-2}$ and $\overline{F_m} \sim m^{-3}$.

On the basis of the foregoing, we remark that similar to the results for the QRW reported in Section~\ref{sec:QRW3}, we see a stark
contrast in the behavior of the survival probability under {\bf Scheme
1} and {\bf Scheme 2} of the measurement dynamics in the case of the TBM.

In Refs.  \cite{Dhar:2015,Dhar:2015-1,Lahiri:2019,Dubey:2021}, study of {\bf Scheme 2} dynamics of the TBM, Eq. (\ref{eq:qua-H}), was carried out for the case where the particle is projected to a state that is in general different from the initial state. More precisely,  in Refs.~\cite{Dhar:2015,Dhar:2015-1}, the initial location of the particle is taken to be site $l$ (thus, the initial state of the particle is $|l\rangle$), while the state to which the instantaneous state of the particle is repeatedly projected to is taken to be $|N\rangle$ (thus, the projection operator is $|N\rangle \langle N|$, and corresponds to measurements being made by a detector placed on site $N$ of the lattice).  An important difference with respect to our work is that in these set of work,  projective measurements were considered to be taking place at regular intervals of length $\tau$, as opposed to stochastic $\tau$ considered in our work.  In these work, it was shown analytically that for large even $N$ and large $m\tau^2/N$, the survival probability $S_m$ decays as $m^{-3/2}$ when the separation between sites $l$ and $N$, measured along the shortest path on the one-dimensional lattice,  is of order $1$ and as $m^{-1/2}$ when the separation is of order $N$. It was also revealed for even $N$ that the survival probability decays to a nonzero constant equal to $1/2$ when the initial location is $l \ne N/2,N$ and to zero for $l=N/2,N$.  In our case of stochastic measurements,  we also observe the $m^{-3/2}$-decay, and the probability decays to zero.  Our results thus serve as an extension of the behavior $\sim m^{-3/2}$ of the survival probability reported in Ref.~\cite{Dhar:2015}
to the case when the measurements are done not at regular but at random intervals distributed according to a wide class of distribution functions:  exponential, power law,  and half-normal for several values of the associated variance. The existence of the scales $m^{\star}_1(N)$ and $m^{\star}_2(N)$ has been discussed in Ref.~\cite{Lahiri:2019} for the case of measurements at regular intervals.

A relevant study of the TBM subject to repeated projective measurements at regular time intervals of length $\tau$, with dynamical evolution following {\bf Scheme 2}, was pursued in Refs. ~\cite{Friedman:2017,Friedman:2017-1}.  In these work, a quantum renewal equation approach was introduced to obtain the first detection probability $F_m$, namely, the probability that a quantum particle that is subject to projective measurements to a given site gets detected at the given site at time $m\tau$ for the first time.  For the TBM defined on the infinite line,  the probability of first detection was shown to exhibit a variety of rich behavior including, e.g., a decay as $m^{-3}$, in stark contrast to the $m^{-3/2}$ decay observed in the related system of classical Brownian motion.  For the same dynamical setup, it was revealed in Ref.~\cite{Thiel:2018} that the first detection depends quite sensitively on the distance between the detector and the initial location of the particle and on $\tau$,  exhibiting scaling and nonanalytical behaviours in appropriate regimes.  In Ref.  \cite{Thiel:2019},  considering the {\bf Scheme 2} of dynamics with measurements at regular intervals of equal length $\tau$,  the authors addressed the influence of the symmetries of a system on the total detection probability $P_{\rm det}$, i.e., the probability to eventually detect the particle after an arbitrary number of detection attempts.  It was demonstrated that this total detection probability is less than unity in symmetric systems, where one may find initial states,  the so-called dark states, that are shielded from the detector by means of destructive interference.  Reference~\cite{Meidan:2019} investigated under {\bf Scheme 2} and considering measurements after every fixed time interval $\tau$ the effect of employing a moving detector on the first detection probability.  For the TBM on the infinite line,  it was shown that the system exhibits a dynamical phase transition at a critical $\tau$,  from a state where the probability of detection decreases exponentially in time and the total detection probability is very small, to one exhibiting a power-law decay and a significantly higher total detection probability.  The average first detected return time, under {\bf Scheme 2} of dynamics with measurements at every fixed interval of length $\tau$,  is known to be quantized, while the return time,  a random variable, is known to exhibit huge fluctuations in appropriate regimes.  In this backdrop,  the work~\cite{Yin:2019} derived explicit expressions for the variance of the return time, together with a classification scheme of the diverging variance based on different underlying physical effects.  Reference~\cite{Thiel:2020} considered {\bf Scheme 2} and measurements after fixed time intervals of length $\tau$ to study the total probability $P_{\rm det}$ in some target state, for example, on the node of a graph after one has made an arbitrary number of detection attempts. An explicit formula for $P_{\rm det}$ in terms of the energy eigenstates of the underlying system was derived,  which was found to be generically $\tau$-independent.  The work is noteworthy in employing the formalism of the paper to obtain a simple upper bound for $P_{\rm det}$.   While the total detection probability $P_{\rm det}$ in classical random walks is unity, the same for a quantum walker, with detection in some state $|d\rangle$,  may for certain initial states $|\psi_{\rm in}\rangle$ take a value smaller than unity. In Refs. \cite{Thiel:2020-1, Thiel:2021}, the authors derived universal bounds  for the quantum total detection probability, under {\bf Scheme 2} of dynamics with measurements after every fixed interval of time equal to $\tau$.  It was shown that the deviation $\Delta P \equiv P_{\rm det}-|\langle \psi_{\rm in}|d\rangle|^2$ satisfies the inequality $\Delta P~{\rm Var}[H]_d \ge |\langle d|[H,D]|\psi_{\rm in}\rangle|^2$, where $D=|d\rangle \langle d|$ is the measurement operator and ${\rm Var}[H]$ measures energy fluctuations in state $|d\rangle$.  Reference~\cite{Liu:2021} studied by considering {\bf Scheme 2} of dynamics the effect of conditional null measurements, done at equal intervals of time of length $\tau$, on a quantum system and displayed a wide variety of rich behavior,  e.g., for systems with built-in symmetry and a degenerate energy spectrum, the null measurements are found to dynamically select the degenerate energy levels, while the non-degenerate levels are effectively wiped out by the measurements.  

In the above backdrop,  we remark that our results reported in this work are complementary to those reported in Refs.~\cite{Dhar:2015,Dhar:2015-1,Friedman:2017,Friedman:2017-1,Thiel:2018,Thiel:2019,Lahiri:2019,Meidan:2019,Yin:2019,Thiel:2020,Thiel:2020-1,Dubey:2021,Thiel:2021,Liu:2021} because of the very different dynamical set-ups (regular $\tau$) considered in these work with respect to ours (stochastic $\tau$). The only exception to the above set of work that considered stochastic $\tau$ in the setting of the TBM with periodic boundary conditions is Ref.~\cite{Kessler:2021}. This work dealt with {\bf Scheme 2} of dynamics, with the gap between consecutive measurements being independent and identically-distributed random variables $\tau$ distributed according to a given distribution $\rho(\tau)$. It was shown that for all $\rho(\tau)$ and finite-dimensional Hamiltonians, the mean detection time is equal to the mean attempt number multiplied by the mean time interval between detection attempts.  Nevertheless,  the issues and results that we report on in this work do not have overlap with this reference either.  Let us remark that for the case of measurements at regular intervals,  our results that in the limit $N \to \infty$, one would observe the behavior $F_m \sim m^{-3}$ is fully consistent with what is reported in Ref.~\cite{Friedman:2017-1}.

\subsubsection{Analytical results}
\label{sec:TBM2b}

For {\bf Scheme 1}, on using the continuous-$\tau$ equivalent of Eqs.~\eqref{eq:qua-Sm-avg-qrw}
and~\eqref{eq:qua-Sm-typical-qrw}, one may obtain 
the average and the typical survival probability for the TBM subject to
projective measurements at random times intervals $\tau$ distributed
according to the exponential and the power-law
distribution, Eqs.~\eqref{eq:qua-exponential-ptau}
and~\eqref{eq:qua-powerlaw-ptau}, respectively. 
One has 
\begin{equation}
\begin{aligned}
        &\overline{S_m}=\exp\left(m\log \int {\rm
        d}\tau~p(\tau)q(\tau)\right),  \\[1ex]
        \label{eq:survival-prob-TBM-projected} 
        &S_m^\star=\exp\left(m \int {\rm
        d}\tau~p(\tau)\log q(\tau)\right). 
\end{aligned}
\end{equation}
In this case,  $q(\tau)$, see Eq. (\ref{eq:qua-qrw-qtau-0}), is nothing
but the probability to be on site $n_0$ after time $\tau$ while starting
from the same site, and is therefore obtained from Eq.~\eqref{eq:qua-pn} as
\begin{align}
q(\tau)= |\psi_{n_0,n_0}(\tau)|^2= \left|\frac{1}{N}\sum_{q=0}^{N-1} \ee^{\ii 2 \gamma \tau \cos (2\pi q/N)}\right|^2.
        \label{eq:qua-tbm-qtau}
\end{align}
The theoretical results so
obtained are compared in Fig.~\ref{fig:qua-tbm-proj} against those obtained in 
numerical implementation of the {\bf Scheme 1} dynamics. We see from the
figure a very good match of the average and the typical survival
probability results. 

For the case $p(\tau)=\delta(\tau-\tau_0)$, we get
$\overline{S_m}=S_m^\star=\exp\left(m \log q(\tau_0)\right)$, and ${\cal
T}=m\tau_0$. In the
limit $\tau_0 \to 0,~m \to \infty$ with ${\cal T}$ kept constant
(frequent measurements at close intervals), using Eq.~\eqref{eq:qua-qrw-qtau-0} and the fact
that $|\psi(0)\ra$ is normalized to unity, it then follows that
$\overline{S_m}=S_m^\star=1-m{\cal O}(\tau_0^2)$. This result implies that to
leading order, there is no evolution of the initial state, an
illustration of the quantum Zeno effect~\cite{Misra:1977}.

For \textbf{Scheme 2}, a semi-analytic approach to obtain
$\ket{\psi_m^{\rm (a)}}$ along the lines employed for the QRW and detailed in
Section~\ref{sec:QRW3b}, involves the following steps:
\begin{enumerate}
        \item For the initial state $|\psi(0)\ra=|n_0\ra$ so that
                $\psi_{n,n_0}(0)=\delta_{n,n_0}$, we have the discrete
                Fourier transform $\widehat{\psi}_{q|n_0}(0)=\exp(-\ii 2\pi
        n_0q/N)$ for $0 \le q \le N-1$.
\item Subsequently, $\widehat{\psi}_{q|n_0}(\tau_1)$, as the outcome of evolution according to~\eqref{eq:qua-psiq-equation} for a random time $\tau_1$ sampled according to either the exponential
distribution~\eqref{eq:qua-exponential-ptau} or the power-law
distribution~\eqref{eq:qua-powerlaw-ptau} and with
                $\widehat{\psi}_{q|n_0}(0)$ as the initial condition, is
                obtained as
                \begin{align}
                \widehat{\psi}_{q|n_0}(\tau_1) = \widehat{\psi}_{q|n_0}(0) \, \ee^{ \ii 2 \gamma \tau_1 \cos(2\pi q/N ) } .
                \end{align}
\item Inverse discrete Fourier transform of the set
$\{\widehat{\psi}_{q|n_0}(\tau_1)\}_{0 \le q \le N-1}$ yields the set
                $\{\psi_{n,n_0}(\tau_1)\}_{0 \le n \le N-1}$. The result
                of a projective measurement at the end of evolution for time
$\tau_1$ to obtain the corresponding leftover component of the
                state is then given by the set
                $\{\psi_{n,n_0}(\tau_1)\}_{0 \le n \le N-1}$ with
                $\psi_{n_0,n_0}(\tau_1)=0$. 
\item We apply steps 1--3 in turn to the leftover component of the 
state corresponding to last projection, to finally obtain the survival probability $S_m$ for a given realization
$\{\tau_\alpha\}_{1\le \alpha \le m}$ of the dynamics.
\end{enumerate}
Figure~\ref{fig:qua-tbm-left} shows a very good agreement for both the
average and the typical survival probability between the numerical
results and those obtained based on the aforementioned semi-analytic
approach.

\section{Conclusions}
\label{sec:conclusions}

In this work, we studied the issue of what happens when a quantum system
undergoing unitary evolution in time is subject to repeated projective measurements to the initial state at
random times. We considered two distinct dynamical scenarios: {\bf
Scheme 1}, in
which the evolution after every projective measurement continues with the
projected component of the instantaneous state, and {\bf Scheme 2}, in which the evolution continues with the
leftover component of the instantaneous state after a 
measurement has been performed.
We focused on a physical quantity of relevance, namely, the survival
probability of the initial state after a certain number $m$ of
measurements have been performed on the system.  Based on results derived
for two representative quantum systems defined on a one-dimensional periodic lattice with a finite number of sites $N$,  (i) the quantum random walk evolving in discrete time
and (ii) the tight-binding model evolving in continuous time, we showed that in {\bf Scheme 1}, both
the average (averaged with respect to different realizations of the set
of random time intervals $\{\tau_\alpha\}_{1\le \alpha \le m}$ between successive
measurements) and the typical survival probability (obtained in a typical
realization of the set $\{\tau_\alpha\}_{1\le \alpha \le m}$) decay as an
exponential in $m$ for large $m$.  One obtains under {\bf Scheme 2} by stark contrast to {\bf Scheme 1} that the behaviour of the survival probability is characterized by two characteristic $m$ values, namely, $m_1^\star(N) \sim N$ and $m_2^\star(N) \sim N^\delta$ with $\delta >1$. These scales are such that (i) for $m$ large and satisfying $m < m_1^\star(N)$, the decay of the survival probability is as $m^{-2}$,  (ii) for $m$ satisfying $m_1^\star(N) \ll m <m_2^\star(N)$, the decay is as $m^{-3/2
        }$,  while (iii) for $m \gg m_2^\star(N)$, the decay is as an exponential.  These results
hold independently of the choice of the distribution of times $\tau_\alpha$. We demonstrate this on the basis of our results obtained for a wide range of distributions including exponential and power-law distributions as well as for the case of measurements at regular intervals.  It would be
interesting to extend our studies to the case of
a many-body quantum system where additional dynamical timescales may
interplay with the average time between successive measurements to
dictate rich static and dynamical behavior. 

\section{Acknowledgements}
The authors acknowledge fruitful discussions with Ashik Iqubal. SG is
grateful to Stefano Gherardini and Haggai Landa for insightful preliminary discussions. SG acknowledges support from the Science and Engineering Research
Board (SERB), India under SERB-TARE scheme Grant No.
TAR/2018/000023, SERB-MATRICS scheme Grant No.
MTR/2019/000560, and SERB-CRG scheme Grant No. CRG/2020/000596. Part of this work was carried out using computational facilities of the Advanced Computing Research Centre,
University of Bristol, UK - \texttt{http://www.bristol.ac.uk/acrc/}.  We thank the anonymous referee for making a number of useful suggestions that helped us improve the presentation significantly.

 

\appendix

\section{Derivation of Eq.~\eqref{eq:qua-Sm-typical-qrw} of the main text}
\label{sec:qua-app1}
In this appendix, we briefly discuss the large deviation (LD) formalism
to obtain Eq.~\eqref{eq:qua-Sm-typical-qrw} of the main text, following Ref.~\cite{Gherardini:2016}. To proceed, let us specialize to the case of $p_\tau$
being a $d$-dimensional Bernoulli distribution. In other words, we
consider the situation in which $\tau$ takes on $d$ possible discrete
values $\tau^{(1)},\tau^{(2)},\ldots,\tau^{(d)}$, with corresponding
probabilities $p^{(1)},p^{(2)},\ldots,p^{(d)}$ satisfying
$\sum_{\alpha=1}^d p^{(\alpha)}=1$. To invoke the LD formalism for the survival probability, consider the
quantity
\begin{align}
        \mathcal{L}(\{\tau_\alpha\}_{1\le \alpha \le m})\equiv\log
        \left(S_m(\{\tau_\alpha\}_{1\le \alpha \le
        m})\right) =\sum_{\alpha=1}^{d}n_{\alpha}\log
        q(\tau^{(\alpha)}),
        \label{eq:qua-Sm-7}
\end{align}
where we have denoted by $n_{\alpha}$ the number of times the value
$\tau^{(\alpha)}$ occurs in the sequence $\{\tau_\alpha\}_{1\le \alpha \le m}$.
The quantity $\mathcal{L}$ is a sum of
i.i.d.~random variables, and its probability distribution is evidently
given by (see Ref.~\cite{Gherardini:2016})
\begin{align}
{\cal P}(\mathcal{L})&=\frac{m!}{n^\prime_{1}!~n^\prime_{2}!~\ldots~
n^\prime_{d}!} ~\prod_{\alpha=1}^{d}
        ~(p^{(\alpha)})^{n^\prime_\alpha},
        \label{eq:qua-Sm-8}
\end{align}
where the quantities
$n^\prime_\alpha$ satisfy the two constraints $\sum\limits_{\alpha=1}^d n^\prime_{\alpha}=m$ and
$\sum\limits_{\alpha=1}^{d}n^\prime_{\alpha}\log
q(\tau^{(\alpha)})=\mathcal{L}$, implying that one has $m \log
q\big(\tau^{(d)}\big)-\mathcal{L}=\sum\limits_{\alpha=1}^{d-1}n_{\alpha}'
~\lambda\big(\tau^{(\alpha)}\big)$, with
$\lambda(\tau^{(\alpha)})\equiv\log q(\tau^{(d)})- \log
q(\tau^{(\alpha)})$. The solution is~\cite{Gherardini:2016}
\begin{align}\label{qua-sol1}
n_{\alpha}'=\frac{m\log q\left(\tau^{(d)}\right)-\mathcal{L}}
{(d-1)~\lambda\left(\tau^{(\alpha)}\right)}; \quad \alpha=1,2,\ldots,d-1,
\end{align}
and $n_d'=m-\sum_{\alpha=1}^{d-1}n_{\alpha}'$. Using this solution in
Eq.~\eqref{eq:qua-Sm-8}, it may be shown that in the limit $m \to \infty$, the
distribution ${\cal P}(\mathcal{L})$ has the following LD form~\cite{Gherardini:2016}
\begin{align}
{\cal P}(\mathcal{L}) \to  {\cal P}({\cal L}/m) \approx\exp \left(
        -mI({\mathcal{L}}/{m}) \right),
\label{qua-app0-eq2}
\end{align}
with 
\begin{align}
&I\left(x\right) \equiv \sum_{\alpha=1}^{d}f(\tau^{(\alpha)})\log\left(\frac{f(\tau^{(\alpha)})}{p^{(\alpha)}}\right), \label{eq:qua-Sm-11}
\\
&f(\tau^{(\alpha)}) \equiv \frac{\log q(\tau^{(d)})-x}{(d-1)\lambda(\tau^{(\alpha)})} \, ; \quad \alpha=1,\ldots,(d-1), \label{qua-appeq2} \\
&f(\tau^{(d)}) \equiv 1-\sum_{\alpha=1}^{d-1}f(\tau^{(\alpha)})  \, .
\end{align}

From Eq.~\eqref{qua-app0-eq2}, it follows that
the minimum of the function $I(\mathcal{L}/m)$ corresponds to the most probable value $\mathcal{L}^{\star}$ of
$\mathcal{L}$ as $m \to \infty$. From Eq.~\eqref{eq:qua-Sm-11}, the condition $\partial
I\left(\mathcal{L}/m\right)/\partial \log
q(\tau^{(\alpha)})|_{\mathcal{L}=\mathcal{L}^\star}=0$ gives, on
performing a series of algebraic manipulations, that~\cite{Gherardini:2016}
\begin{align}
\mathcal{L}^\star=m\sum_{\alpha=1}^{d} p^{(\alpha)}\log
q(\tau^{(\alpha)}).
\label{eq:qua-Sm-15}
\end{align}

Using Eqs.~\eqref{eq:qua-Sm-7} and~\eqref{qua-app0-eq2}, one may obtain
an LD form for the
distribution of the survival probability $S_m$ as~\cite{Gherardini:2016}
\begin{align}
{\cal P}(S_m)\approx\exp\left(-m ~J(S_m)\right),
        \label{eq:qua-Sm-16}
\end{align}
with $J(S_m)\equiv{\rm min}_{\mathcal{L}:\mathcal{L}=\log
S_m}I(\mathcal{L}/m)$. The value $S_m^\star$ that minimizes the function $J(S_m)$ is the most probable value of the
survival probability in the limit $m \to \infty$; one gets~\cite{Gherardini:2016}
\begin{align}
S_m^{\star}=\exp\left(m\sum_{\alpha=1}^{d} ~p^{(\alpha)}\log
        q(\tau^{(\alpha)})\right),
        \label{eq:qua-Sm-18}
\end{align}
which rewritten suitably yields Eq.~\eqref{eq:qua-Sm-typical-qrw} of the main text. 



\begin{thebibliography}{99}
\bibitem{qua-note-hbar} In this work, we take the Planck's constant to be unity.
\bibitem{ptau}
  We use the notation $p(\tau)$ (respectively, $p_\tau$)
  to denote the distribution for the case when $\tau$ is a
  continuous  (respectively, a discrete) variable.
\end{thebibliography}
\end{document}